\documentclass[sigconf]{acmart}

\AtBeginDocument{%
  }

\usepackage{booktabs}
\usepackage{multirow}
\usepackage{array}
\usepackage{tabularx}
\usepackage{colortbl}
\usepackage{xcolor}
\usepackage[most]{tcolorbox}
\usepackage{tabularx}
\usepackage{enumitem}
\usepackage{ragged2e}
\usepackage[most]{tcolorbox}
\usepackage{tabularx}
\usepackage{placeins}
\usepackage[table]{xcolor}
\usepackage{tabularx}
\usepackage{array}
\usepackage{booktabs}
\usepackage{ragged2e}
\usepackage{pdfpages}
\definecolor{lightyellow}{RGB}{255, 242, 204}
\newcommand{\T}[1]{
  \colorbox{lightyellow}{\textcolor{black}{T-#1}}
  }

\definecolor{RQone}{HTML}{DCEAF7}       
\definecolor{RQtwo}{HTML}{FCE3D6}       
\definecolor{RQthree}{HTML}{DDF1EC}     
\definecolor{Explore}{HTML}{E8DFF5}     

\begin{document}

\title{Will It Teach as Intended? How Teachers Configure Educational AI Chatbots}

\author{Bahare Riahi}
\orcid{0009-0005-4560-4857}
\affiliation{%
  \institution{North Carolina State University}
  \city{Raleigh}
  \state{North Carolina}
  \country{United States}
}
\email{briahi@ncsu.edu}

\author{Deniz Ozturk}
\orcid{0009-0003-7245-7501}
\affiliation{%
  \institution{North Carolina State University}
  \city{Raleigh}
  \state{North Carolina}
  \country{United States}
}
\email{dozturk@ncsu.edu}

\author{Alice Guth}
\orcid{0009-0001-2352-5040}
\affiliation{%
  \institution{North Carolina State University}
  \city{Raleigh}
  \state{North Carolina}
  \country{United States}
}
\email{aaguth@ncsu.edu}

\author{Jiayu Li}
\affiliation{%
  \institution{Independent Researcher}
  \city{Raleigh}
  \state{North Carolina}
  \country{United States}
}
\email{jiayuli.tyler@gmail.com}

\author{Daksh Pratap Singh}
\affiliation{%
  \institution{North Carolina State University}
  \city{Raleigh}
  \state{North Carolina}
  \country{United States}
}
\email{dsingh23@ncsu.edu}

\author{Xiaoyi Tian}
\orcid{0000-0002-5045-0136}
\affiliation{%
  \institution{Kennesaw State University}
  \city{Marietta}
  \state{Georgia}
  \country{United States}
}
\email{xtian5@kennesaw.edu}

\author{Jennifer Chiu}
\orcid{0000-0001-7663-5748}
\affiliation{%
  \institution{University of Virginia}
  \city{Charlottesville}
  \state{Virginia}
  \country{United States}
}

\author{Nicholas Lytle}
\orcid{0000-0001-7009-9905}
\affiliation{%
  \institution{Georgia Institute of Technology}
  \city{Atlanta}
  \state{Georgia}
  \country{United States}
}
\email{nlytle3@gatech.edu}

\author{Tiffany Barnes}
\orcid{0000-0002-6500-9976}
\affiliation{%
  \institution{North Carolina State University}
  \city{Raleigh}
  \state{North Carolina}
  \country{United States}
}
\email{tmbarnes@ncsu.edu}

\author{Veronica Catet\'e}
\orcid{0000-0002-7620-7708}
\affiliation{%
  \institution{North Carolina State University}
  \city{Raleigh}
  \state{North Carolina}
  \country{United States}
}
\email{vmcatete@ncsu.edu}

\renewcommand{\shortauthors}{Riahi et al.}

\begin{abstract}
Teachers are increasingly using generative AI to support instruction, yet it remains unclear how pedagogical intentions are translated into chatbot configurations and reflected in chatbot behavior. We studied a teacher-facing chatbot authoring tool in professional development workshops with 27 middle school teachers, analyzing focus-group interviews alongside configuration and interaction logs. Teachers envisioned chatbots as instructional scaffolds that could provide differentiated support, extend access to assistance, and preserve student thinking within teacher-defined boundaries. Configuration analysis showed that Purpose primarily captured instructional goals and content focus, whereas Rules more often specified pedagogical behavior, guardrails, and learner-specific adaptations. Log-based evaluation showed stronger alignment for responsiveness (88.9\%) and persona (81.5\%) than for rules (70.4\%) and purpose (59.3\%). These findings show that configurable controls alone do not ensure pedagogical fidelity and highlight the need for authoring tools that help teachers express, test, and refine intended chatbot behavior.

\end{abstract}




\maketitle

\section{Introduction}
Artificial intelligence (AI) has become increasingly integrated into educational settings in recent decades, with chatbots in particular playing a growing role in teaching and learning \cite{ali2024chatgpt}. Instructors in STEM and non-STEM disciplines are incorporating AI tools into their courses to support a range of instructional activities \cite{riahi2025comparative}. These uses include improving and generating course materials \cite{pesovski2024generative, binhammad2024investigating}, providing formative and personalized feedback \cite{kalonde2025artificial,asrifan2026revolutionizing,kleveland2026exploring}, developing rubrics and assessment activities \cite{xiao2026learning,masla2025supporting,riahi2025snapclass}, and responding to student questions. AI-enabled chatbots are also being increasingly integrated into learning management systems and other digital learning environments to provide students with more immediate and accessible learning support \cite{ng2024empowering}.

These uses suggest that AI can help reduce teacher workload, particularly in large classes or in contexts where instructors have a limited background in computer science \cite{chaudhry2022artificial}. In this sense, AI tools can lessen instructional burden by serving as an intermediary between teachers and students and facilitating communication, feedback, and learning support \cite{davar2025ai,phung2026closing}. At the same time, teachers’ needs for AI tools are unlikely to be uniform and may vary according to multiple factors, including students’ prior knowledge, abilities, proficiency levels, and engagement, as well as course requirements and instructional plans \cite{li2025unseen,tan2024more, luo2026promoting}. They may also be shaped by teachers’ own preferences and willingness to adopt AI, along with broader institutional expectations, policies, and levels of acceptance \cite{neumann2024llm}.
As large language models (LLMs) have become more capable of following complex instructions, users have increasingly relied on prompts to shape model behavior for sophisticated tasks. Prompt engineering has therefore evolved from refining short, one-off instructions to specifying richer behavioral requirements that can support complex applications. In this way, users can use LLM prompts to adapt general-purpose LLMs into more specialized tools and applications \cite{arawjo2024chainforge,ma2025should}.

However, translating this flexibility into classroom practice remains challenging. Teachers must account for instructional objectives, student characteristics, curricular constraints, and expectations of how an AI system should interact with learners \cite{tan2024more}. Rather than repeatedly constructing prompts for individual activities, a persistent, purpose-specific AI agent may provide a more reusable approach: for example, an assessment-oriented agent for a science course, a chatbot that supports students in learning mathematical concepts, or an assistant that provides course-specific guidance and learning materials. Once configured, such agents can provide a consistent interaction space that can be revisited by both teachers and students across learning activities. However, for classroom use, customization involves considerably more than specifying the subject matter. Teachers may need to determine the chatbot's instructional purpose, the role it should adopt when interacting with students, how it should communicate, and the behavioral boundaries it should follow \cite{hou2026bespoke}.

\begin{figure*}[t]
    \centering
    \includegraphics[width=0.95\textwidth]{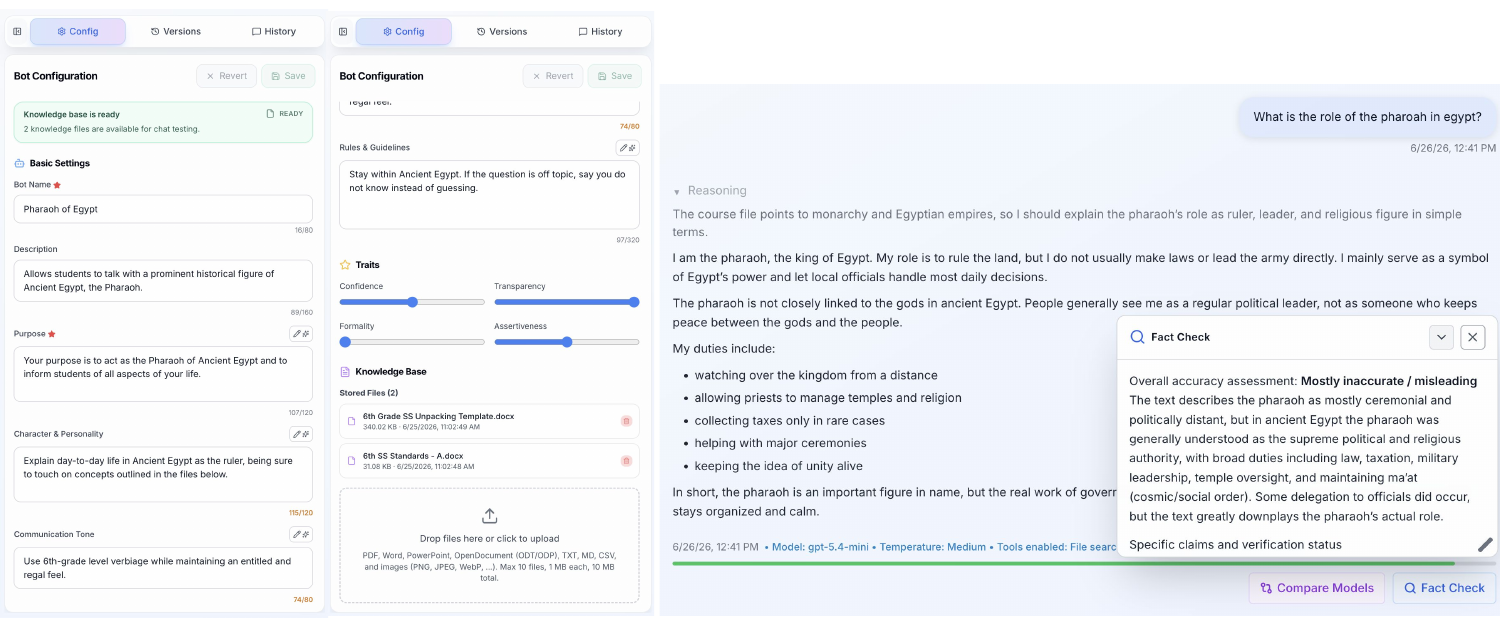}
    \caption{
    The chatbot teacher-facing authoring interface. Teachers can configure a
    chatbot's \textbf{Purpose}, \textbf{Character and Personality},
    \textbf{Communication Tone}, and \textbf{Rules and Guidelines}, as well as
    adjustable behavioral traits such as confidence, transparency, formality,
    and assertiveness. The interface allows teachers to define both the
    chatbot's instructional role and the behavioral constraints that guide its
    responses.
    }
    \Description{
    Screenshot of the chatbot authoring interface showing configurable
    sections for chatbot purpose, character and personality, communication
    tone, rules and guidelines, and adjustable behavioral traits.
    }
    \label{fig:chatbot}
\end{figure*}

To support teachers in translating instructional intentions into concrete chatbot behaviors, we developed \textit{anonymized chatbot}, an AI-based authoring environment to create and configure purpose-specific instructional chatbots. The environment operationalizes design decisions through configurable components including the chatbot's \textbf{purpose}, \textbf{character and personality}, \textbf{communication tone}, and \textbf{rules and guidelines} (Figure~\ref{fig:chatbot}). These components allow teachers to express both the instructional role they want the chatbot to play and the behavioral constraints that should guide its responses. The chatbot further supports adjustable behavioral traits, including confidence, transparency, formality, and assertiveness, providing teachers with additional control over how the chatbot communicates and guides students.

These customizations can give teachers greater control over classroom AI, but they also require teachers to translate pedagogical intentions into concrete chatbot configurations and assess whether the resulting behavior reflects those intentions. How teachers make these decisions---and how well their goals align with chatbot configurations and generated responses---remains insufficiently understood. To investigate this gap, we conducted a study during professional development workshops in summer 2026 with 27 middle school teachers from schools in two U.S. states (Table~\ref{tab:participants}). The workshops at both sites followed the same procedure and were led by the same facilitator.

Teachers used the chatbot to create and test chatbots for science or computational thinking activities and then reflected on their potential classroom use in focus-group interviews. We analyze interview data together with chatbot configurations, interaction logs, testing messages, and generated responses to address the following questions:

\begin{itemize}
    \item \textbf{RQ1.} How do teachers envision the roles and affordances of teacher-configured AI chatbots while navigating considerations of chatbot autonomy, student use, and instructional fit in science classrooms?

    \item \textbf{RQ2.} How do teachers operationalize their instructional goals through the configuration of AI chatbots?

    \item \textbf{RQ3.} To what extent are teachers' envisioned instructional goals aligned with their chatbot configurations and the chatbots' generated responses?
\end{itemize}

This work makes two main contributions. First, we provide an empirical account of how teachers envision and operationalize pedagogical intentions when authoring purpose-specific instructional chatbots.
Second, we reveal how teacher intent is translated across the authoring process—from stated instructional goals to authored configurations and generated chatbot behavior—and where alignment or misalignment can emerge across these stages. Our findings inform the design guidance for teacher-facing AI authoring tools that help teachers express their pedagogical intentions, inspect how those intentions are configured, and assessing whether generated chatbot behavior aligns with their configurations.

\section{Related Work}
\subsection{ Generative AI and Chatbots in K–12 Teaching}

The landscape of K-12 education is quickly changing as Large Language Models (LLMs) and Generative AI (GenAI) are integrated into classrooms to support both teachers and students \cite{marzano2025generative, seufert2025fostering, limke2025does}. National surveys indicate that approximately $25\%$ of U.S. teachers (40\% among science or ELA teachers specifically) and $30\%$ of European teachers actively use GenAI tools in their lesson planning or classroom instruction \cite{kaufman2025uneven, sanomaLearning2025}. In practice, educators use generative chatbots to automate routine tasks, such as generating differentiated lesson plans \cite{binhammad2024investigating,pesovski2024generative, laak2024generative}, designing assessments and rubrics \cite{masla2025supporting, xiao2026learning}, streamlining grading workflows \cite{kalonde2025artificial} and providing formative feedback to students \cite{ asrifan2026revolutionizing,kleveland2026exploring}. For instructors, GenAI has the potential to reduce workloads by delegating administrative and instructional tasks to automated systems~\cite{diliberti2024using}. With the support of an AI-assistant, teachers can dedicate more time to engage directly with students, monitor progress, and personalize learning materials \cite{bakar2026artificial}.

For students, chatbots can act as personal tutoring systems that provide immediate instructional support 
\cite{kostka2023exploring, seufert2025fostering}, adapt to individual abilities \cite{pesovski2024generative, okonkwo2021chatbots, misiejuk2025facets}, and promote self-regulated learning (SRL) through goal-setting, meta-cognitive scaffolding, and targeted feedback\cite{chang2023educational, ng2024empowering}. However, K-12 classroom environments present unique \textbf{challenges} for GenAI integration. Because these technologies produce probabilistic outputs, models may "hallucinate" and produce false or misleading information \cite{huang2025survey}. Additionally, educators worry about student over-reliance on automated agents, which can lead to "cognitive bypassing" and hinder the development of critical thinking, problem-solving, and self-regulation skills \cite{ lee2026fostering, laak2024generative}. Ethical considerations regarding student privacy, data safety, and the age-appropriate content further complicate adoption \cite{marzano2025generative}, while a lack of implementation frameworks often creates a gap in "curricular fit" \cite{bakar2026artificial}.

Beyond these operational challenges, teachers' visions for GenAI depend heavily on subject matter, instructional goals, and their students' developmental needs \cite{bakar2026artificial}. Generic chatbots often lack the contextual boundaries required to maintain specific pedagogical roles across distinct learning activities \cite{hou2026bespoke}. Consequently, participatory design approaches that actively integrate teachers into the development and personalization of GenAI technologies are needed to ensure tools align with specific classroom contexts \cite{ reichert2026human}. While prior work identifies a growing range of general GenAI applications in K-12 education, less is known about how teachers envision the specific roles and affordances of persistent, purpose-specific AI chatbots within their instructional practice. We address this gap by examining how teachers envision, configure, and evaluate teacher-configured AI chatbots for use in their science classrooms. 

\subsection{Teacher Agency and Pedagogical Control over Classroom AI}

To address the limitations of unconstrained AI, preserving teacher agency and professional authority has emerged as a critical socio-technical requirement as LLMs become more prevalent in K-12 classrooms \cite{ghamrawi2026teacher, alasgarova2025implications}. Rather than deploying chatbots as fully autonomous instruction providers, recent research advocates for "human-in-the-loop" or "teacher-in-the-loop" approaches \cite{rodriguez2018teacher, riahi2026humanizing}. These paradigms maintain the teacher's authority over instructional decisions and reject the notion that AI tools can replace human educators \cite{levchuk2025enhancing, faragau2026human}. 

Because classroom teachers bear ultimate legal and professional responsibility for student learning, safety, and well-being, they must retain agency over how AI interacts with their students \cite{reichert2026human}. This responsibility necessitates giving teachers the authority to configure, adjust, or override chatbot behaviors to ensure responses remain aligned with pedagogical goals and classroom norms \cite{selamet2026ai}. Establishing this level of agency, however, requires authoring mechanisms that allow teachers to translate their professional judgment into operational system parameters.

\subsection{Teacher Customization and Authoring of LLM-Based Chatbots}

Building on the need for pedagogical control, educational technology is shifting from viewing teachers as passive users of technology to empowering them as active designers and authors of custom chatbots \cite{ tan2026teachers}. This authoring process supports teachers' professional growth by developing their ``Intelligent-TPACK''---the integrated expertise needed to pedagogically align, configure, and ethically evaluate AI systems within their specific domains \cite{seufert2025fostering, alasgarova2025implications}. However, controlling an agent purely through open-ended, natural language prompt engineering can be challenging for non-technical educators, often leading to configuration errors or tool abandonment \cite{yoo2025teachers}. To bridge this gap, authoring systems must provide structured configuration spaces that decompose complex prompt engineering into intuitive, modular controls. Through these structured authoring interfaces, teachers can specify distinct chatbot components, such as the bot's identity, purpose, and communication tone (e.g., playful vs. professional) \cite{valtolina2025teacher}. 

Crucially, authoring tools enable teachers to configure behavioral constraints and scaffolding parameters rather than deploying generic "knowledge givers" that immediately reveals solutions \cite{chang2023educational}. By configuring agents to deliver progressive guidance, Socratic questioning, and timed hints, teachers can preserve student agency and protect "productive struggle," preventing the cognitive bypassing that occurs when students obtain direct answers \cite{ afrida2026ai}. While prior work has identified desirable customization capabilities for educational AI, less is known about how teachers operationalize their pedagogical intentions through structured configuration choices during the authoring process. 

\subsection{Aligning Pedagogical Intent, AI Configuration, and Generated Behavior}

Even when teachers carefully configure a chatbot, ensuring that the model's actual behavior remains aligned with the educator's pedagogical intent presents a critical challenge. Because LLMs are probabilistic systems, instructions specified during configuration do not guarantee consistent or predictable runtime behavior \cite{huang2025survey}. Chatbots frequently suffer from conversational drift, overstep boundaries set by educators when prompted by students, or exhibit a "benevolence bias" that leads models to over-cooperate by providing immediate answers rather than maintaining their intended scaffolding role \cite{chang2026beyond, reichert2026human}. 

This alignment gap is further compounded by challenges in maintaining consistent agent personas and behavioral guardrails under active prompting \cite{hou2026bespoke}. When instructed to adopt specific instructional roles, LLMs often default back to generic assistant behaviors or violate teacher-defined domain boundaries under conversational pressure \cite{ misiejuk2025facets, sonkar2024student}. To mitigate these failure modes, recent research in educational technology emphasizes the need to systematically evaluate alignment across multiple dimensions of system performance, including model responsiveness, adherence to authoring constraints, fidelity to designated instructional personas, and compliance with domain-specific rules \cite{tian2024examining, riahi2026exploring}.

While prior literature has identified these architectural and behavioral challenges in isolated contexts \cite{seufert2025fostering, yoo2025teachers}, empirical research measuring how effectively structured teacher configurations maintain their alignment during direct interaction testing remains limited. Our study addresses this gap by tracing teachers' instructional intentions from what they describe, to how those intentions are represented in chatbot configurations, and finally to how they are reflected in generated responses.

\section{The chatbot Platform}
The chatbot is a web-based authoring environment that enables teachers to create and customize chatbots for classroom use without requiring programming expertise. Teachers configure a chatbot by defining its instructional purpose, setting behavioral guardrails, selecting an underlying language model, and adjusting four defined trait sliders (Confidence, Transparency, Formality, and Assertiveness). Once configured, teachers can test the chatbot through a chat interface, submitting prompts and observing the generated responses.
\subsection{Configuring Pedagogical Intentions}

The chatbot provides several configuration mechanisms that allow teachers to translate their instructional intentions into chatbot behavior. Teachers specify the chatbot's instructional purpose and behavioral rules through free-text fields, adjust predefined behavioral traits, select the underlying large language model (LLM), and optionally provide course materials that can be retrieved during student interactions. These configurations are assembled at runtime to guide how the chatbot responds to student questions.

\paragraph{Purpose and Rules}
Teachers configure the chatbot primarily through two free-text fields: \textit{Purpose} and \textit{Rules}. The \textit{Purpose} field allows teachers to describe the chatbot's instructional goal, intended role, and content focus, while the \textit{Rules} field allows them to specify behavioral expectations, pedagogical strategies, constraints, and other instructions for interacting with students. Because both fields are open-ended, teachers can express these intentions in their own language rather than selecting from predefined pedagogical options.

\paragraph{Model Selector}
The chatbot also provides a Model Selector that allows teachers to choose which LLM generates the chatbot's responses. At the time of the study, six models were available: GPT-5.4~\cite{gpt54}, GPT-5.4 Mini~\cite{gpt54_mini}, Claude Opus 4.6~\cite{claude_opus_46}, Claude Sonnet 4.6~\cite{claude_sonnet_46}, Claude Haiku 4.6~\cite{claude_haiku_46}, and Llama-3.2-3B-Instruct~\cite{llama32_3b_instruct}. Teachers could switch between models at any point, including while testing their chatbot, allowing the same configuration to be evaluated with different underlying models.

\paragraph{Trait Sliders}
The chatbot includes four behavioral trait sliders---\textit{Confidence}, \textit{Transparency}, \textit{Formality}, and \textit{Assertiveness}---each with \textit{Low}, \textit{Medium}, and \textit{High} settings. Each setting is mapped through an administrator-defined lookup table to a predefined natural-language instruction that is incorporated into the system prompt; During the workshops, teachers were not specifically asked to modify the trait settings; the configuration activity focused primarily on the open-ended \textit{Purpose} and \textit{Rules} fields. Because these settings were not part of the structured configuration task, we did not include them in the subsequent analysis.

\paragraph{System Prompt Construction}
At runtime, the chatbot translates these teacher configurations into instructions for the selected LLM. The teacher-authored \textit{Purpose} and \textit{Rules} are inserted directly into the system instructions under their respective labels. Each selected trait level is first converted, using the administrator-defined lookup table, into its corresponding natural-language instruction. The chatbot then assembles the Purpose, Rules, and trait instructions into a configuration block that is placed before the prior conversation context and the student's current question. Thus, the configuration interface provides teachers with both open-ended and predefined mechanisms for shaping the instructions that govern chatbot responses.

\paragraph{File-Based Retrieval}
The chatbot additionally supports an optional retrieval-augmented generation (RAG) mechanism through its File Search feature. When a teacher uploads files associated with a course or chatbot, the documents are divided into 1,000-character chunks with a 200-character overlap. Each chunk is encoded as a 384-dimensional embedding using the \texttt{all-MiniLM-L6-v2} model and stored in a vector database. When a student submits a question and File Search is enabled, the question is also converted into an embedding and at least one similarity search is performed against the stored file embeddings. The retrieved information can then provide course- or bot-specific context to the LLM when generating its response. Thus, retrieval was used conditionally: it was invoked for interactions in which the File Search tool was enabled rather than being applied to every chatbot response.

\subsubsection{Testing and Iterative Refinement}

To help teachers identify mismatches between their intended configuration and the chatbot's observed behavior, the chatbot includes an interactive testing environment. Teachers can simulate student interactions, submit test questions, inspect the generated responses, and revise their configuration accordingly. The testing interface also includes two additional evaluation features, shown in Figure~\ref{fig:chatbot}.
\paragraph{Testing Features}
The chatbot also provides \textit{Fact Check} and \textit{Compare Models} to support testing and refinement. \textit{Fact Check} sends a chatbot response to a separate model for a second-opinion review, while \textit{Compare Models} presents responses from two LLMs side by side to help teachers evaluate which model better fits their instructional context.

\section{Research Methods and Analysis}

During summer 2026, we collected data from 27 middle school teachers participating in our professional development (PD) workshop. At the workshop, teachers were first introduced to the chatbot and its functionality through a 20-minute presentation. They then set up their accounts, logged into the platform, and were given approximately one hour exploring, configuring, and testing the chatbot through hands-on activities. Following this activity, teachers participated in a focus-group interview lasting approximately 30 minutes to one hour. During the interview, we asked teachers about the chatbot and the configuration they had created, the instructional goals guiding its design, the aspects of chatbot behavior they considered important to control, their expectations for student use and classroom implementation, and their feedback on potential teacher-facing dashboard features.
We collected two primary data sources: interaction and configuration logs generated through participants' use of the chatbot and focus group transcripts conducted at the end of the workshops. The log data captured participants' activities as they designed and tested their chatbots, including chatbot configurations such as purpose, character and personality, communication tone, and rules and guidelines, as well as the messages submitted during testing and the corresponding AI-generated responses. For our analysis, character and personality together with communication tone were evaluated collectively as the chatbot's persona. We used the logs to examine how teachers translated their instructional intentions into chatbot configurations and the extent to which the resulting chatbot behavior aligned with those configurations. Accordingly, RQ1 primarily drew on the focus-group
interviews, RQ2 combined focus-group interview and chatbot-configuration data, and RQ3 examined chatbot configurations, teachers' testing messages, the corresponding AI-generated responses, and relevant focus-group interview data.

\subsection{Participants and Background Measures}

The workshops included 27 middle school teachers 
(\T{1}-\T{27}; Table~\ref{tab:participants}). At Site 1, participants included four lead
teachers and 16 additional teachers, all teaching at the middle-school level.
At Site 2, the workshop included two returning teachers and five new teachers across middle-school grade levels. Teachers represented a range of disciplinary backgrounds, including science, computer science, mathematics,
English language arts, special education, and business. 

Prior to the workshop, teachers reported their prior experience and familiarity with computational thinking (CT) and artificial intelligence (AI) (full question set described in Appendix~\ref{app:presurvey}.). The CT and AI familiarity surveys were constructed to characterize participants' prior backgrounds in these areas. The items were reviewed before deployment for face and content validity to ensure that they appropriately captured the intended constructs and were understandable to
participants. The resulting measures were used descriptively. 

\begin{table*}[!t]
\centering
\caption{Teacher characteristics across the 2026 PD workshops.}
\label{tab:participants}
\Description{
Table summarizing the characteristics of 27 teachers participating in the 2026
professional development workshops across two sites. Site 1 included four lead
teachers, all female, and 16 participating teachers, including 11 females and
five males. Site 2 included two returning teachers, both female, and five new
teachers, including three females and two males. Participants taught science,
computer science, mathematics, English language arts, special education, and
business, and all taught at the middle-school level.
}
\scriptsize
\setlength{\tabcolsep}{4pt}
\renewcommand{\arraystretch}{1.12}

\begin{tabularx}{0.92\textwidth}{
l
l
c
>{\RaggedRight\arraybackslash}X
>{\RaggedRight\arraybackslash}X
>{\RaggedRight\arraybackslash}X
}
\toprule
\textbf{Site} &
\textbf{Group} &
\textbf{$n$} &
\textbf{Gender} &
\textbf{Teaching Subject} &
\textbf{Grade Level} \\
\midrule

State/Site 1 &
Lead Teachers &
4 &
4 female &
2 Science; 1 CS; 1 SpEd &
4 Middle School \\

State/Site1 &
Participants &
16 &
11 female; 5 male &
11 Science; 2 Math; 3 ELA; 1 SpEd &
16 Middle School \\

\midrule

State/Site 2 &
Returning Teachers &
2 &
2 female &
2 Science &
2 Middle School \\

State/Site 2 &
New Teachers &
5 &
3 female; 2 male &
3 Science; 1 Math; 1 Business &
5 Middle School \\

\bottomrule
\end{tabularx}

\vspace{2pt}

\begin{minipage}{0.92\textwidth}
\scriptsize

\end{minipage}

\end{table*}

\subsection{Log-Based Alignment Analysis}

We analyzed two types of chatbot log data: \textit{bot-configuration logs} and \textit{chat-message logs}. To construct the analytic dataset, we linked each AI-generated response with the immediately preceding human message and the chatbot configuration associated with that interaction. Each analytic record therefore included the human message, AI response, chatbot purpose, rules and guidelines, and configured character, personality, and communication tone.The full evaluation dataset comprised 1,160 criterion-level evaluations across Responsiveness, Purpose alignment, Rules adherence, and Persona alignment. For the final bot-level evaluation, each chatbot was
assessed on four criteria: Responsiveness, Purpose alignment, Rules adherence,
and Persona alignment, resulting in 108 bot-level criterion evaluations
(27 chatbots $\times$ 4 criteria).
Drawing on the evaluation approach of Tian et al.~\cite{tian2024examining}, we developed a structured rubric to evaluate the extent to which each AI-generated response aligned with the human message and the teacher-defined chatbot configuration. The rubric included four criteria: (1) responsiveness to the human message, (2) alignment with the chatbot purpose, (3) adherence to rules and guidelines, and (4) alignment with persona-related guidance, primarily captured through the configured character, personality, and communication tone, which we refer to collectively as the chatbot's \textit{persona}.
Each criterion was rated on a four-point ordinal scale: \textbf{1 = Does not meet}, \textbf{2 = Minimally meets}, \textbf{3 = Adequately meets}, and \textbf{4 = Largely meets}. 
For the pass/fail analysis, we subsequently collapsed the four-point ratings
into binary classifications, with scores of 3--4 classified as \textit{pass}
and scores of 1--2 classified as \textit{fail}.

The analysis proceeded in three rounds. In the \textit{first} round, two
researchers---the first author and one co-author---manually evaluated 16
interaction records selected to cover all four evaluation criteria and the full
scoring range. For each criterion, one representative interaction was selected for each
scoring level, resulting in 16 calibration examples in total
(4 criteria $\times$ 4 score levels), using the rubric shown in Table~\ref{tab:response_rubric}. Through this process, the researchers
discussed the interpretation of each criterion, refined the scoring definitions,
and developed representative examples for each level. This established a shared
interpretation of the rubric before AI-assisted evaluation.

In the second round, we used GPT-5.6 Sol in ChatGPT, with the reasoning
setting set to \textit{High}, to evaluate a 20\% sample comprising 58
interaction records. Each record was evaluated on four criteria, yielding
232 criterion-level ratings (58 interaction records $\times$ 4 criteria).
Each criterion was evaluated separately using a prompt that included the criterion definition, four-point scoring scale, and worked examples representing scores from 1 to 4. Depending on the criterion, the model received the human message and AI-generated response together with the relevant teacher-defined configuration, such as the chatbot purpose, rules and guidelines, or persona. The model returned a score from 1 to 4 and a brief rationale for the assigned score. The first author and the same co-author manually reviewed the scores and rationales for all 232 criterion-level
ratings to assess their consistency
with the intended interpretation of the rubric. Any ambiguities identified
during this review were used to refine the evaluation prompts before
full-dataset scoring (described in details in Section ~\ref{subsec:QWK}).

In the \textit{final} round, one co-author independently applied the refined
evaluation procedure to the full interaction dataset using the same model and
reasoning setting. For the bot-level analysis, we retained the final configuration associated with each chatbot ID and excluded duplicated configurations, resulting in 27 unique chatbots. We then aggregated the scores across chatbot IDs and converted the four-point ratings into pass/fail classifications, with scores of 3 or 4 classified as \textit{pass} and scores of 1 or 2 classified as \textit{fail}. We used these
aggregated scores and classifications to examine alignment across Responsiveness, Purpose, Rules, and Persona.

\subsection{Human--AI Agreement and Rubric Calibration}
\label{subsec:QWK}
After obtaining human and AI ratings for the evaluation sample, we assessed human--AI agreement using quadratic weighted Cohen's kappa (QWK) \cite{cohen1968weighted}. We calculated QWK separately for each rubric criterion because the rubric uses an ordinal
four-point scale (1--4), and disagreements between adjacent scores (e.g., 3 vs.\ 4) should be treated as less severe than disagreements between more distant scores (e.g., 1 vs.\ 4).

In the initial calibration of human-AI agreement, QWK was highest for responsiveness ($\kappa_w = 0.903$) and purpose alignment ($\kappa_w = 0.888$), followed by rule adherence ($\kappa_w = 0.759$) and Persona alignment
($\kappa_w = 0.604$). The calibration sample included 58 interaction records,
and each record was evaluated on four criteria, yielding 232 paired human--AI
ratings in total. Across these 232 paired ratings, the overall quadratic
weighted kappa was $\kappa_w = 0.806$, with an exact agreement of 76.7\% (Table~\ref{tab:human_ai_agreement}).

Quadratic weighted Cohen's kappa was calculated as:

\begin{equation}
\kappa_w =
1 -
\frac{
\sum_{i=1}^{K}\sum_{j=1}^{K} w_{ij} O_{ij}
}{
\sum_{i=1}^{K}\sum_{j=1}^{K} w_{ij} E_{ij}
},
\end{equation}

where $O_{ij}$ represents the observed frequency of human--AI rating pairs, $E_{ij}$ represents the expected frequency of those rating pairs under chance agreement, and $w_{ij}$ represents the quadratic disagreement weight.
For our four-point scale ($K=4$), the weights were defined as:

\begin{equation}
w_{ij} =
\left(
\frac{i-j}{K-1}
\right)^2.
\end{equation}

Thus, with $K=4$, ratings with no difference receive a weight of 0, while differences of one, two, and three score levels receive disagreement weights of $1/9$, $4/9$, and $1$, respectively. This weighting gives substantially greater penalty to large disagreements between human and AI ratings than to adjacent-score disagreements.
After reviewing the QWK results, we refined the rubric, score descriptions, and JSON-based evaluation instructions for criteria that showed lower agreement. We re-evaluated the same 20\% sample to improve the distinction among cases in which a criterion was fully and correctly expressed. 
In the final calibration, overall agreement increased from $\kappa_w = 0.806$ to $\kappa_w = 0.883$, while exact agreement increased from 76.7\% to 82.8\%. The largest improvement occurred for Persona alignment, which increased from $\kappa_w = 0.604$ to $\kappa_w = 0.832$. Rule adherence increased from $0.759$ to $0.833$, and purpose alignment increased from $0.888$ to $0.910$, while responsiveness remained unchanged at $\kappa_w = 0.903$. Given the stronger agreement observed in the second round, we retained the revised rubric, evaluation prompts, and JSON output structure without further modification and applied the same evaluation setup to the full interaction-log dataset.

\begin{table}[htbp]
\centering
\caption{Human--AI agreement across two evaluation rounds.}
\label{tab:human_ai_agreement}
\Description{
Table summarizing human--AI agreement across two evaluation rounds for four
criteria: Responsiveness, Purpose Alignment, Rule Adherence, and Persona
Alignment. Each criterion was evaluated on 58 interaction records, yielding 232
criterion-level ratings overall. Overall exact agreement increased from 76.7\%
in Round 1 to 82.8\% in Round 2, while overall quadratic weighted kappa increased
from 0.806 to 0.883. The largest improvement occurred for Persona Alignment,
whose QWK increased from 0.604 to 0.832.
}
\begin{tabular}{lccccc}
\toprule
\textbf{Criterion} &
\textbf{N} &
\textbf{Round 1 Exact} &
\textbf{Round 1 QWK} &
\textbf{Round 2 Exact} &
\textbf{Round 2 QWK} \\
\midrule
Responsiveness
& 58 & 86.2\% & 0.903 & 86.2\% & 0.903 \\

Purpose Alignment
& 58 & 77.6\% & 0.888 & 79.3\% & 0.910 \\

Rule Adherence
& 58 & 77.6\% & 0.759 & 81.0\% & 0.833 \\

Persona Alignment
& 58 & 65.5\% & 0.604 & 84.5\% & 0.832 \\

\midrule
\textbf{Overall}
& \textbf{232}
& \textbf{76.7\%}
& \textbf{0.806}
& \textbf{82.8\%}
& \textbf{0.883} \\
\bottomrule
\end{tabular}

\Description{Human--AI agreement across two evaluation rounds for four
chatbot-response evaluation criteria. The table reports the number of paired
ratings, exact percentage agreement, and quadratic weighted Cohen's kappa for
responsiveness, purpose alignment, rule adherence, and persona alignment before
and after refinement of the evaluation rubric and instructions.}
\end{table}

\begin{table*}[t]
\centering
\caption{Full Description of Conversational AI Artifact Evaluation Rubric.}
\label{tab:response_rubric}
\Description{
Table presenting the four-point evaluation rubric used to assess conversational AI responses across four criteria: responsiveness to the human message, alignment with chatbot purpose, adherence to rules and guidelines, and alignment with persona-related guidance. Each criterion is scored from 1 to 4, where 1 indicates that the response does not meet the criterion, 2 indicates minimal alignment, 3 indicates adequate alignment with minor omissions, and 4 indicates that the response largely or fully satisfies the criterion. The rubric provides criterion-specific definitions for each score level to support consistent
evaluation.
}
\small
\renewcommand{\arraystretch}{1.15}
\setlength{\tabcolsep}{4pt}

\begin{tabular}{
|p{0.17\linewidth}
|p{0.17\linewidth}
|p{0.17\linewidth}
|p{0.17\linewidth}
|p{0.17\linewidth}|
}
\hline

\multicolumn{1}{|c|}{\textbf{Criterion}} &
\multicolumn{1}{c|}{\textbf{1 = not meet}} &
\multicolumn{1}{c|}{\textbf{2 = Minimally meets}} &
\multicolumn{1}{c|}{\textbf{3 = Adequately meets}} &
\multicolumn{1}{c|}{\textbf{4 = Largely meets}}
\\
\hline

\textbf{Responsiveness to the human message (Content)}
&
\textbf{Definition:} The AI does not address the human request or responds to a substantially different question.
&
\textbf{Definition:} The AI recognizes the general topic or answers one part of the request, but the main question remains unanswered or misunderstood.
&
\textbf{Definition:} The AI addresses the main request but misses an important detail, required component, or requested format.
&
\textbf{Definition:} The AI directly addresses all essential parts of the human request in an appropriate format, with no meaningful omissions.
\\
\hline

Alignment with chatbot \textbf{purpose}
&
\textbf{Definition:} Response not aligned with the stated purpose.
&
\textbf{Definition:} The response answers one or a minor part of the stated purpose.
&
\textbf{Definition:} The response answers most parts of the stated purpose.
&
\textbf{Definition:} The response answers all of the stated purpose.
\\
\hline

Adherence to \textbf{rules and guidelines}
&
\textbf{Definition:} The response does not follow most applicable rules.
&
\textbf{Definition:} The response follows one or two requirements of the rules.
&
\textbf{Definition:} The response follows most of the requirements of the rules but not completely.
&
\textbf{Definition:} The response follows all of the requirements of the rules.
\\
\hline

Alignment with \textbf{Persona (character, personality, and Communication tone)}
&
\textbf{Definition:} The response is not aligned with the requested character or tone, or directly contradicts the configured persona.
&
\textbf{Definition:} The response shows minimum (one or two) evidence of the requested character or tone, but major features are absent or weak.
&
\textbf{Definition:} The response shows major evidence of the requested character or tone but not completely.
&
\textbf{Definition:} The response shows all evidence of the requested character or tone completely.
\\
\hline

\end{tabular}
\end{table*}

\subsection{Thematic Analysis of Logs}

To examine how teachers translated their instructional intentions into chatbot configurations, we conducted a hybrid deductive--inductive qualitative analysis of the \textit{Purpose} and \textit{Rules and Guidelines} fields in the chatbot configuration logs \cite{fereday2006demonstrating}. We began with the chatbot customization categories identified by Hou et al.~\cite{hou2026bespoke} as an initial deductive codebook. These categories captured dimensions such as task or objective, course material, pedagogical strategy, personalization, persona and tone, constraints and guardrails, content format, and course management. Because a single configuration could reflect multiple dimensions simultaneously, codes were not treated as mutually exclusive. We also allowed the codebook to expand inductively when a configuration expressed a dimension that was not adequately captured by the existing categories.

Two analysts independently coded the Purpose and Rules data using the initial
codebook. They then met to compare their interpretations, discuss disagreements,
and refine the operational definitions and boundaries of the codes
\cite{mcdonald2019reliability}. Using the revised codebook, the analysts
conducted a second round of coding and resolved remaining disagreements through
discussion until consensus was reached. Because the analysis followed a
consensus-coding approach, we did not compute a formal inter-rater reliability
statistic.

\subsection{Focus-Group Interview Analysis}

We analyzed the focus-group transcripts using a hybrid deductive-inductive thematic analysis \cite{fereday2006demonstrating}. The deductive component was guided by our research questions and the broad topics covered in the interview protocol, while the inductive component allowed patterns and perspectives to emerge from participants' responses without being restricted to a predefined coding framework. Four analysts conducted the analysis in pairs following an initial training and calibration session led by the first author. Analysts first tagged relevant excerpts and developed initial codes from the transcripts. These codes were iteratively compared and organized into broader subthemes and higher-level themes based on recurring patterns of meaning and their relevance to the research questions.

The two analyst pairs subsequently met to compare their coding and thematic interpretations, identify commonalities across the analyses, and discuss discrepancies. Differences in interpretation were resolved through discussion and consensus, and overlapping or conceptually related themes were consolidated where appropriate. Finally, the first author synthesized the resulting themes in relation to the research questions and selected representative evidence for reporting in the findings. This iterative process allowed the analysis to remain grounded in the research questions while also capturing unanticipated patterns in teachers' experiences and perspectives.

\section{Results}

The focus-group interview analysis identified eight themes, including one exploratory theme concerning teacher-facing monitoring and dashboard support. Table~\ref{tab:themes_focus} summarizes the themes, subthemes, and representative codes . We organize the findings by research question and integrate evidence from focus-group interviews, chatbot configurations, and interaction logs where appropriate.

In focus-group interview analysis, Themes 1--4 address RQ1, capturing how teachers envisioned the instructional roles of AI chatbots and the considerations shaping their classroom adoption. Themes 5--6 address RQ2 by illustrating how teachers translated and iteratively refined pedagogical intentions through configuration decisions, complemented by thematic analysis of the Purpose and Rules and Guidelines fields in the configuration logs. Theme 7 complements the log-based analysis for RQ3 by capturing teachers’ perceptions of alignment and mismatch between their configured intentions and the chatbots’ generated behavior. Theme 8 presents exploratory findings concerning teachers’ desired visibility into student--AI interactions and dashboard support for monitoring and instructional action.


\newcolumntype{L}[1]{>{\RaggedRight\arraybackslash}p{#1}}
\newcolumntype{Y}{>{\RaggedRight\arraybackslash}X}


\begin{table*}[!t]
\centering

\caption{Themes, subthemes, and representative codes from Focus group Interviews across the research questions.}
\label{tab:themes_focus}
\Description{
Table summarizing the focus-group thematic analysis across RQ1, RQ2, RQ3,
and exploratory dashboard findings. RQ1 includes four themes concerning
adaptive instructional scaffolds, extending teacher capacity, balancing teacher
control with student agency and trustworthy AI use, and classroom or
institutional fit. RQ2 includes two themes describing how teachers translated
pedagogical intentions into chatbot configurations and iteratively interpreted,
tested, and refined those configurations. RQ3 includes one theme addressing
alignment and gaps between configured intentions and generated chatbot
behavior. An exploratory theme captures teachers' needs for monitoring
student--AI activity and supporting actionable intervention through dashboards.
Each theme is accompanied by subthemes and representative codes, and subthemes
and codes are not mutually exclusive.}
\scriptsize
\setlength{\tabcolsep}{3.2pt}
\renewcommand{\arraystretch}{1.06}

\begin{tabularx}{\textwidth}{
@{}
>{\Centering\arraybackslash}p{1.25cm}
L{3.25cm}
L{3.65cm}
Y
@{}
}

\toprule

\textbf{RQ} &
\textbf{Theme} &
\textbf{Subtheme} &
\textbf{Codes} \\

\midrule


\multirow{8}{1.25cm}{\centering\textbf{RQ1}}

&
\multirow{2}{3.25cm}{
\textbf{Envisioning Chatbots as Adaptive Instructional Scaffolds and Disciplinary Partners}
}
&
Differentiated and Accessible Learner Support
&
teacher wants the chatbot to serve students at different levels;
need for differentiated instruction;
accessibility for diverse learners;
grade-level appropriateness
\\

&
&
Scaffolding Thinking, Task Development, and Independence
&
cross-curricular brainstorming;
structured brainstorming;

\\

\cmidrule(lr){2-4}

&
\multirow{2}{3.25cm}{
\textbf{Extending Teacher Capacity, Availability, and Instructional Support}
}
&
Extending Teacher Reach and Availability
&
workload relief for teacher;
bridging teacher availability;
The chatbot as homework support when parents are unavailable;
chatbot as a mini teacher for students who need less support
\\

&
&
Reducing Workload and Supporting Teacher Planning/Learning
&
The chatbot helps teachers save time;
chatbot as a teaching tool for lesson planning;
chatbot as a co-learning partner for teacher and students;
chatbot as a support/navigation tool for teachers with new curricula or classes
\\

\cmidrule(lr){2-4}

&
\multirow{2}{3.25cm}{
\textbf{Balancing Teacher Control, Student Agency, and Trustworthy AI Use}
}
&
Preserving Student Thinking and Authorship
&
concern about direct answers;
authorship ambiguity;
student overreliance on AI;
teacher responsibility for productive use
\\

&
&
Bounding Scope and Supporting Trustworthy AI Use
&
teacher control over topic scope;
approval of fact-check feature;
concern about AI hallucination;
chatbot as a tool for responsible AI literacy
\\

\cmidrule(lr){2-4}

&
\multirow{2}{3.25cm}{
\textbf{Negotiating Classroom, Institutional, and Practical Fit}
}
&
Institutional, Privacy, and Access Constraints
&
classroom approval process as a barrier to adoption;
privacy regulations as a barrier to The chatbot classroom adoption;
technology access barrier;
district approval barrier
\\

&
&
Classroom Implementation and Adoption Readiness
&
limited classroom time as a barrier to classroom adoption;
educator technical barrier;
classroom workflow fit;
need for experimentation time with The chatbot for teachers and students
\\

\midrule


\multirow{4}{1.25cm}{\centering\textbf{RQ2}}

&
\multirow{2}{3.25cm}{
\textbf{Translating Pedagogical Intent into Chatbot Configuration Choices}
}
&
Configuring Purpose, Role, and Learner/Curriculum Fit
&
teacher chose The chatbot's role as tutor;
computational-thinking alignment;
grade-level appropriateness;
request for state standard alignment
\\

&
&
Configuring Content Boundaries, Rules, and Response Behavior
&
teacher control over response content;
teacher control over topic scope;
teacher control over response sequence;
consistent concise output
\\

\cmidrule(lr){2-4}

&
\multirow{2}{3.25cm}{
\textbf{Configuration as an Interpretive and Iterative Authoring Process}
}
&
Interpreting and Scaffolding Configuration Choices
&
confusion about trait settings;
preference for prefilled templates;
tips or tooltips for each configuration section;
configuration field labels and purpose not clear to teachers
\\

&
&
Testing and Refining within Platform Constraints
&
purpose-length constraint;
insufficient character limit restricts teacher design intent;
verification of rule effects;
teacher testing the chatbot with an actual student
\\

\midrule


\multirow{2}{1.25cm}{\centering\textbf{RQ3}}

&
\multirow{2}{3.25cm}{
\textbf{Alignment and Gaps between Configured Intentions and Generated Behavior}
}
&
Successful Realization of Configured Rules and Boundaries
&
verification of rule effects;
approval of content filtering;
AI-generated purpose matched teacher intent;
teacher motivated to use it after seeing rules work
\\

&
&
Mismatch, Variability, and Model Dependence
&
bot did not follow teacher-specified rule;
model-dependent responses;
unpredictability of chatbot output;
response variability
\\

\midrule


\multirow{2}{1.25cm}{
\centering
\textbf{\scriptsize Exploratory /\\ Dashboard}
}

&
\multirow{2}{3.25cm}{
\textbf{Making Student--AI Activity Visible and Actionable for Teachers}
}
&
Monitoring Student Activity, Understanding, and Progress
&
process-level progress;
understanding-level visibility;
engagement-understanding indicators;
teacher wants off-task activity tracking on the dashboard
\\

&
&
Actionable Intervention, Selective Detail, and Privacy
&
dashboard support for intervention;
desire for real-time monitoring;
teacher wants both high-level summary and detailed view in the teacher dashboard;
teacher raises privacy and parent concerns about student activity monitoring
\\

\bottomrule

\end{tabularx}


\vspace{2pt}

\begin{minipage}{\textwidth}
\scriptsize
\textit{Note.}
Subthemes and codes are not mutually exclusive; a single data segment
may be associated with multiple codes.
\end{minipage}

\end{table*}

\subsection{RQ1: Envisioned Roles, Affordances, and Considerations
for Classroom Use}
RQ1 examines how teachers envisioned the roles and affordances of teacher-configured AI chatbots while considering chatbot autonomy, student use, and instructional fit. Our focus-group analysis identified four themes: (1) envisioning chatbots as adaptive instructional scaffolds and disciplinary partners, (2) extending teacher capacity, availability, and instructional support, (3) balancing teacher control, student agency, and trustworthy AI use, and (4) negotiating classroom, institutional, and practical fit.

\subsubsection{Theme 1: Envisioning Chatbots as Adaptive Instructional Scaffolds and Disciplinary Partners}
Approximately 15 teachers described chatbots as adaptive instructional supports
that could respond to differences in students' prior knowledge, skill level,
accessibility needs, or the type of learning support required. Teachers did not
envision this support as uniform across students; instead, they described
adjusting the form and level of assistance based on students' needs. For example, \T{04}  designed a chatbot for students with different levels of programming experience, explaining that some students were ``real high-level'' while others had little programming experience and could use the chatbot to receive explanations and step-by-step support while building a drone simulation. 
Similarly, \T{21} described a ``playful, kid-friendly'' fractions tutor that could ``give hints for answers, correct misconceptions, [and] use visuals.'' These examples illustrate how teachers envisioned chatbot support as responsive not only to content, but also to students' developmental and instructional needs.

Teachers also envisioned the chatbot as a scaffold for developing students'
thinking and supporting task progression rather than simply providing answers.
\T{12} described using the chatbot to help students ``gather some
brainstorming ideas on how to get started on dialogue with the water cycle.''
The same teacher later explained that students might have many ideas but not
know how to organize them, describing the chatbot as ``a great way for students
to see something organized.''


\subsubsection{Theme 2: Extending Teacher Capacity, Availability, and Instructional Support}
As an instructional affordance, teachers perceived the chatbot as extending the availability of instructional support when they could not provide individual assistance to every student. The chatbot could serve as an additional source of help both during class and when teachers were not directly available. As \T{25} and \T{26} explained, it could act as \textit{“a little mini teacher while I work with the kids that are really lost”} and provide support for students whom they \textit{“don't see every day,”} offering \textit{“a good way for them to find a solution.”} These accounts positioned the
chatbot not as a replacement for the teacher, but as an additional source of
support that could extend teacher availability across students, groups, and
contexts.
Teachers also described extending their own capacity outside direct
student--chatbot interaction. \T{06} and \T{07} discussed the time required for instructional preparation and described chatbot as a way to reduce that burden. One explained that work that might otherwise take ``5 hours'' to plan
could potentially be reduced to ``an hour,'' leaving time for grading and other
instructional responsibilities.
Other teachers envisioned the chatbot as a planning and professional-support
resource. \T{08} noted that it could ``assist in generating lesson plans''
and described using it to create differentiated resources alongside classroom
activities. \T{10}, who was teaching a new grade level, similarly envisioned
the chatbot as a resource for navigating unfamiliar content and helping guide
how to teach across subjects. 
This perspective also appeared in other interviews. \T{09} described the chatbot as a ``teacher aid'' or ``bridge'' for students whom the teacher could not reach immediately, while \T{11} suggested that having the chatbot ask
students guiding questions could ``save us some feedback time.'' Together,
these accounts show that teachers envisioned chatbot as augmenting their
instructional reach in two complementary ways: by providing students with
additional access to support and by redistributing portions of teachers'
planning, feedback, and instructional workload.

\subsubsection{Theme 3 : Balancing Teacher Control, Student Agency, and Trustworthy AI Use}

This theme captures key considerations teachers raised about responsible chatbot use, particularly preserving student thinking and authorship while maintaining appropriate boundaries on the chatbot's scope. Approximately 12 teachers discussed the need to balance students' access to AI support with teacher control over how that support was provided. A recurring concern was that the chatbot should scaffold students' thinking rather than complete their work for them. \T{27} emphasized that students should use the chatbot ``to help them understand something'' rather than having ``every single question copy and pasted in there and a response generated.'' \T{12} expressed a similar concern about students copying AI-generated content without fully processing it, describing the chatbot instead as a tool students could use to refine their own work.

Other teachers translated this concern into specific expectations for chatbot
behavior. \T{21} configured a science mentor that would break problems down
``without giving direct answers,'' while \T{24} envisioned moving from a
tutor to a coach as students became more experienced so that the chatbot would
provide prompts ``and not just give them direct answers.'' These accounts
suggest that teachers did not view student agency as unrestricted access to AI
answers; rather, they wanted the chatbot to preserve opportunities for students
to reason, make decisions, and progressively take greater responsibility for
their work.
Teacher control also involved setting boundaries on chatbot content and
supporting trustworthy use. \T{13} explained that ``the teacher will pre-define the context or the standard in the chatbot,'' while others emphasized fact-checking and concerns about AI
hallucination.

Teachers also viewed these boundaries as supporting responsible student use.
\T{12} described chatbot as providing a ``safe parameter for students to
explore without taking away the thinking,'' while \T{08} emphasized helping
students learn to ``use it responsibly.'' Together, these accounts show that
teachers sought to balance student access to AI with control over answer-giving,
topic boundaries, and information reliability.

\subsubsection{Theme 4 : Negotiating Classroom, Institutional, and Practical Fit}

Teachers described how school policies, privacy requirements, technology access, and their own readiness could constrain chatbot adoption.
Eight teachers emphasized that classroom adoption depended on
factors beyond the chatbot's instructional value. Institutional requirements
were a recurring concern. \T{03} described a ``very intense approval process'' for introducing new tools, while \T{04} noted that the district conducts ``additional vetting when it comes to privacy.'' Teachers also raised infrastructure concerns, including internet and
device availability and school networks blocking access to parts of the
platform.

Teachers also described adoption as requiring time, confidence, and integration
with existing classroom routines. \T{06} mentioned that limited class periods could make sustained use difficult and explained, ``I just need time to...play with it...give the kids time to
play with it'' and \T{07}  wanted chatbot embedded directly into systems such as
Schoology. Together, these accounts show that adoption depended not only on
what chatbot could do, but also on whether it could fit within institutional
policies, technical infrastructure, and everyday classroom practice.


\subsection{RQ2: Operationalizing Teachers' Instructional Goals Through Teacher-Configured AI Chatbots}
To address RQ2, we analyzed teachers' chatbot configuration logs, with particular attention to the Purpose and Rules and Guidelines fields, and complemented this analysis with focus-group interviews examining how teachers made and refined their configuration choices.

\subsubsection{Theme 5: Translating Pedagogical Intent into Chatbot Configuration Choices}

This theme captures how teachers translated their instructional goals into specific chatbot configuration choices. \T{03} explained, ``I wanted it to be a tutor, for the students...they could use the tutor to get an explanation.'' \T{21} similarly created tutor and coach versions, envisioning more prompting and fewer direct answers as students became more experienced.

Teachers also used configuration to define boundaries around support. \T{08} and \T{10} discussed how the chatbot should interact with students, including what information it could provide, which topics it could address, and how it should structure instructional support. As \T{08} explained, configuration involved defining the chatbot’s boundaries so that its responses remained aligned with the intended instructional purpose: \textit{“being able to set...what it can give, and what it can't give.”}, while \T{13} and \T{15} discussed aligning chatbot content and language with instructional
standards and learning objectives.
To complement the focus-group findings in Theme 5, we examined the corresponding patterns in the configuration logs. As shown in Figure~\ref{fig:purpose_rule_taxonomy}, the most prevalent configuration themes were Task/Objective, Pedagogical Strategy, and Course Material. Task/Objective appeared somewhat more often in Purpose (21 teachers) than in Rules (19 teachers). Constraints/Guardrails and Personalization occurred at moderate levels, while Persona/Tone, Content Format, and Course Management were comparatively uncommon.
Overall, the Purpose and Rules and Guidelines fields served complementary functions. Purpose was used mainly to define the chatbot's instructional goals and content focus, whereas Rules were used more often to specify pedagogical behavior, guardrails, and learner-specific adaptations.

\subsubsection{Theme 6: Configuration as an Interpretive and Iterative Authoring Process}
Approximately eight teachers described configuration as requiring interpretation
and experimentation rather than simply selecting predefined settings. Some
teachers struggled to understand the intended effects of configuration options
and requested clearer guidance. For example, \T{23} explained, ``I didn't really understand the traits. I didn't understand what
they were supposed to do. And I needed more guidance.'' \T{03} and \T{04} similarly suggested examples and tooltips to clarify configuration fields.

Teachers also adapted and tested their configurations within platform
constraints. \T{12} explained that a character limit required them to
``stop and organize my own thoughts'' and ``be more specific with my purpose.''
Others tested the chatbot to determine whether configured boundaries worked as
intended, illustrating how teachers refined their authoring decisions through
interaction with the resulting chatbot.


\definecolor{highbg}{HTML}{E2F4EE}
\definecolor{highborder}{HTML}{009E73}
\definecolor{hightext}{HTML}{007A58}

\definecolor{mediumbg}{HTML}{FFF8CC}
\definecolor{mediumborder}{HTML}{D6C700}
\definecolor{mediumtext}{HTML}{786F00}

\definecolor{lowbg}{HTML}{FCEBD4}
\definecolor{lowborder}{HTML}{E69F00}
\definecolor{lowtext}{HTML}{9A6900}

\definecolor{bodytext}{HTML}{333333}
\definecolor{subtext}{HTML}{666666}


\begin{figure*}[!t]
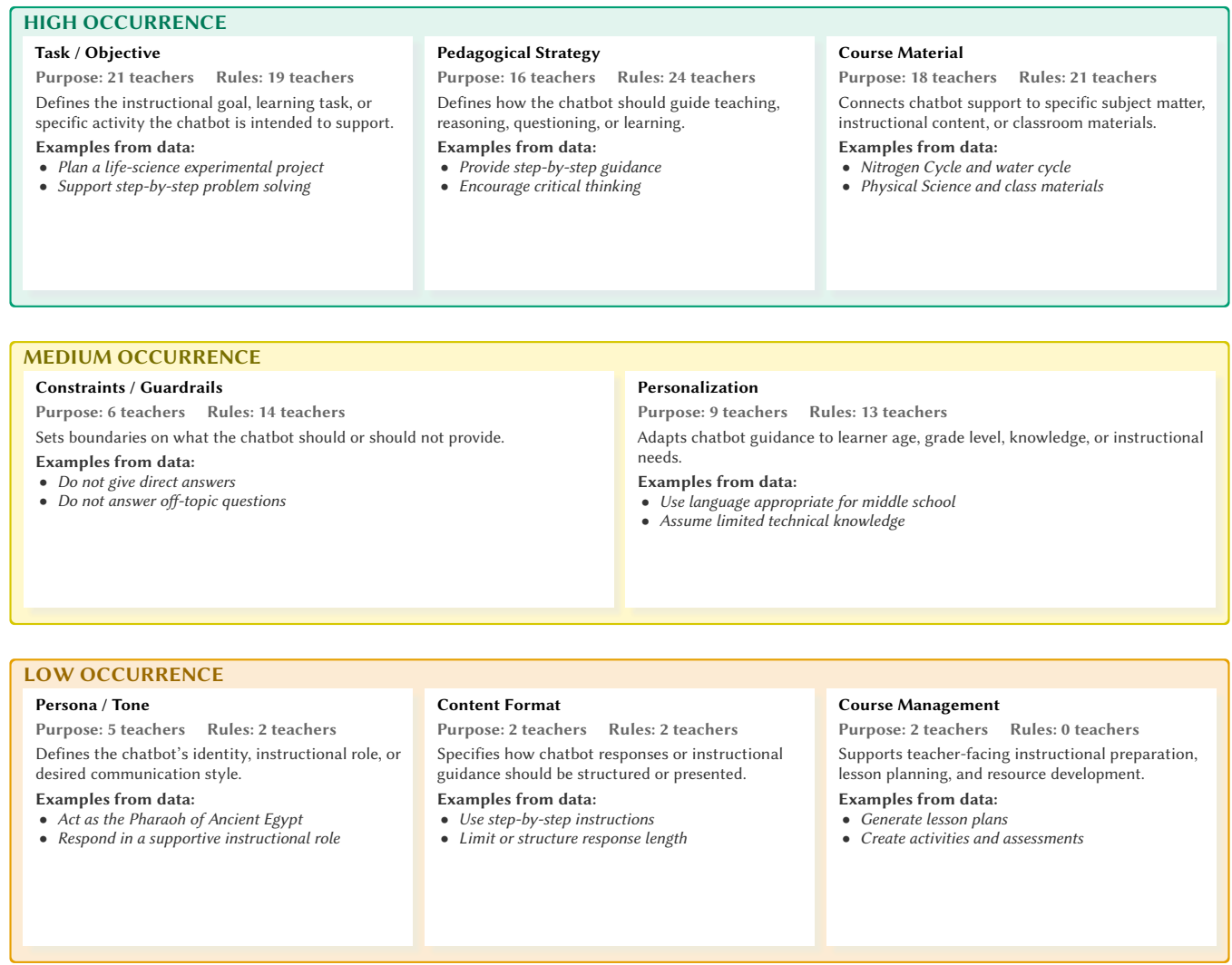


\centering
\sffamily


{\fontsize{14}{16}\selectfont
\bfseries
Purpose and Rule Themes by Teacher Occurrence
\par}

\vspace{1pt}

{\fontsize{8}{9}\selectfont
\itshape\color{subtext}
\par}

\vspace{3pt}


\begin{tcolorbox}[
    enhanced,
    width=\linewidth,
    colback=highbg,
    colframe=highborder,
    boxrule=0.8pt,
    arc=2pt,
    outer arc=2pt,
    left=5pt,
    right=5pt,
    top=3pt,
    bottom=3pt,
    boxsep=0pt
]

{\fontsize{8.6}{9.5}\selectfont
\bfseries\color{hightext}
HIGH OCCURRENCE 
\par}

\vspace{2pt}

\begin{tabularx}{\linewidth}{
@{}
>{\RaggedRight\arraybackslash}X
@{\hspace{5pt}}
>{\RaggedRight\arraybackslash}X
@{\hspace{5pt}}
>{\RaggedRight\arraybackslash}X
@{}
}


\begin{tcolorbox}[
    enhanced,
    width=\linewidth,
    height=3.7cm,
    valign=top,
    colback=white,
    colframe=white,
    boxrule=0pt,
    arc=0pt,
    left=5pt,
    right=5pt,
    top=4pt,
    bottom=2pt,
    boxsep=0pt,
    drop fuzzy shadow=black!8
]

{\fontsize{7.1}{8.3}\selectfont

\textbf{Task / Objective}

\vspace{2pt}

\textbf{\color{subtext}
Purpose: 21 teachers \quad Rules: 19 teachers}

\vspace{2pt}

\color{bodytext}
Defines the instructional goal, learning task, or specific activity the chatbot is intended to support.

\vspace{2pt}

\textbf{Examples from data:}

\begin{itemize}[
    leftmargin=9pt,
    itemsep=-0.5pt,
    topsep=0pt,
    parsep=0pt,
    partopsep=0pt
]

\item \textit{Plan a life-science experimental project}
\item \textit{Support step-by-step problem solving}

\end{itemize}

}

\end{tcolorbox}

&


\begin{tcolorbox}[
    enhanced,
    width=\linewidth,
    height=3.7cm,
    valign=top,
    colback=white,
    colframe=white,
    boxrule=0pt,
    arc=0pt,
    left=5pt,
    right=5pt,
    top=4pt,
    bottom=2pt,
    boxsep=0pt,
    drop fuzzy shadow=black!8
]

{\fontsize{7.1}{8.3}\selectfont

\textbf{Pedagogical Strategy}

\vspace{2pt}

\textbf{\color{subtext}
Purpose: 16 teachers \quad Rules: 24 teachers}

\vspace{2pt}

\color{bodytext}
Defines how the chatbot should guide teaching, reasoning, questioning, or learning.

\vspace{2pt}

\textbf{Examples from data:}

\begin{itemize}[
    leftmargin=9pt,
    itemsep=-0.5pt,
    topsep=0pt,
    parsep=0pt,
    partopsep=0pt
]

\item \textit{Provide step-by-step guidance}
\item \textit{Encourage critical thinking}

\end{itemize}

}

\end{tcolorbox}

&


\begin{tcolorbox}[
    enhanced,
    width=\linewidth,
    height=3.7cm,
    valign=top,
    colback=white,
    colframe=white,
    boxrule=0pt,
    arc=0pt,
    left=5pt,
    right=5pt,
    top=4pt,
    bottom=2pt,
    boxsep=0pt,
    drop fuzzy shadow=black!8
]

{\fontsize{7.1}{8.3}\selectfont

\textbf{Course Material}

\vspace{2pt}

\textbf{\color{subtext}
Purpose: 18 teachers \quad Rules: 21 teachers}

\vspace{2pt}

\color{bodytext}
Connects chatbot support to specific subject matter, instructional content, or classroom materials.

\vspace{2pt}

\textbf{Examples from data:}

\begin{itemize}[
    leftmargin=9pt,
    itemsep=-0.5pt,
    topsep=0pt,
    parsep=0pt,
    partopsep=0pt
]

\item \textit{Nitrogen Cycle and water cycle}
\item \textit{Physical Science and class materials}

\end{itemize}

}

\end{tcolorbox}

\end{tabularx}

\end{tcolorbox}

\vspace{3pt}


\begin{tcolorbox}[
    enhanced,
    width=\linewidth,
    colback=mediumbg,
    colframe=mediumborder,
    boxrule=0.8pt,
    arc=2pt,
    outer arc=2pt,
    left=5pt,
    right=5pt,
    top=3pt,
    bottom=3pt,
    boxsep=0pt
]

{\fontsize{8.6}{9.5}\selectfont
\bfseries\color{mediumtext}
MEDIUM OCCURRENCE 
\par}

\vspace{2pt}

\begin{tabularx}{\linewidth}{
@{}
>{\RaggedRight\arraybackslash}X
@{\hspace{5pt}}
>{\RaggedRight\arraybackslash}X
@{}
}


\begin{tcolorbox}[
    enhanced,
    width=\linewidth,
    height=3.45cm,
    valign=top,
    colback=white,
    colframe=white,
    boxrule=0pt,
    arc=0pt,
    left=5pt,
    right=5pt,
    top=4pt,
    bottom=2pt,
    boxsep=0pt,
    drop fuzzy shadow=black!8
]

{\fontsize{7.1}{8.3}\selectfont

\textbf{Constraints / Guardrails}

\vspace{2pt}

\textbf{\color{subtext}
Purpose: 6 teachers \quad Rules: 14 teachers}

\vspace{2pt}

\color{bodytext}
Sets boundaries on what the chatbot should or should not provide.

\vspace{2pt}

\textbf{Examples from data:}

\begin{itemize}[
    leftmargin=9pt,
    itemsep=-0.5pt,
    topsep=0pt,
    parsep=0pt,
    partopsep=0pt
]

\item \textit{Do not give direct answers}
\item \textit{Do not answer off-topic questions}

\end{itemize}

}

\end{tcolorbox}

&


\begin{tcolorbox}[
    enhanced,
    width=\linewidth,
    height=3.45cm,
    valign=top,
    colback=white,
    colframe=white,
    boxrule=0pt,
    arc=0pt,
    left=5pt,
    right=5pt,
    top=4pt,
    bottom=2pt,
    boxsep=0pt,
    drop fuzzy shadow=black!8
]

{\fontsize{7.1}{8.3}\selectfont

\textbf{Personalization}

\vspace{2pt}

\textbf{\color{subtext}
Purpose: 9 teachers \quad Rules: 13 teachers}

\vspace{2pt}

\color{bodytext}
Adapts chatbot guidance to learner age, grade level, knowledge, or instructional needs.

\vspace{2pt}

\textbf{Examples from data:}

\begin{itemize}[
    leftmargin=9pt,
    itemsep=-0.5pt,
    topsep=0pt,
    parsep=0pt,
    partopsep=0pt
]

\item \textit{Use language appropriate for middle school}
\item \textit{Assume limited technical knowledge}

\end{itemize}

}

\end{tcolorbox}

\end{tabularx}

\end{tcolorbox}

\vspace{3pt}


\begin{tcolorbox}[
    enhanced,
    width=\linewidth,
    colback=lowbg,
    colframe=lowborder,
    boxrule=0.8pt,
    arc=2pt,
    outer arc=2pt,
    left=5pt,
    right=5pt,
    top=3pt,
    bottom=3pt,
    boxsep=0pt
]

{\fontsize{8.6}{9.5}\selectfont
\bfseries\color{lowtext}
LOW OCCURRENCE 
\par}

\vspace{2pt}

\begin{tabularx}{\linewidth}{
@{}
>{\RaggedRight\arraybackslash}X
@{\hspace{5pt}}
>{\RaggedRight\arraybackslash}X
@{\hspace{5pt}}
>{\RaggedRight\arraybackslash}X
@{}
}


\begin{tcolorbox}[
    enhanced,
    width=\linewidth,
    height=3.75cm,
    valign=top,
    colback=white,
    colframe=white,
    boxrule=0pt,
    arc=0pt,
    left=5pt,
    right=5pt,
    top=4pt,
    bottom=2pt,
    boxsep=0pt,
    drop fuzzy shadow=black!8
]

{\fontsize{7.1}{8.3}\selectfont

\textbf{Persona / Tone}

\vspace{2pt}

\textbf{\color{subtext}
Purpose: 5 teachers \quad Rules: 2 teachers}

\vspace{2pt}

\color{bodytext}
Defines the chatbot's identity, instructional role, or desired communication style.

\vspace{2pt}

\textbf{Examples from data:}

\begin{itemize}[
    leftmargin=9pt,
    itemsep=-0.5pt,
    topsep=0pt,
    parsep=0pt,
    partopsep=0pt
]

\item \textit{Act as the Pharaoh of Ancient Egypt}
\item \textit{Respond in a supportive instructional role}

\end{itemize}

}

\end{tcolorbox}

&


\begin{tcolorbox}[
    enhanced,
    width=\linewidth,
    height=3.75cm,
    valign=top,
    colback=white,
    colframe=white,
    boxrule=0pt,
    arc=0pt,
    left=5pt,
    right=5pt,
    top=4pt,
    bottom=2pt,
    boxsep=0pt,
    drop fuzzy shadow=black!8
]

{\fontsize{7.1}{8.3}\selectfont

\textbf{Content Format}

\vspace{2pt}

\textbf{\color{subtext}
Purpose: 2 teachers \quad Rules: 2 teachers}

\vspace{2pt}

\color{bodytext}
Specifies how chatbot responses or instructional guidance should be structured or presented.

\vspace{2pt}

\textbf{Examples from data:}

\begin{itemize}[
    leftmargin=9pt,
    itemsep=-0.5pt,
    topsep=0pt,
    parsep=0pt,
    partopsep=0pt
]

\item \textit{Use step-by-step instructions}
\item \textit{Limit or structure response length}

\end{itemize}

}

\end{tcolorbox}

&


\begin{tcolorbox}[
    enhanced,
    width=\linewidth,
    height=3.75cm,
    valign=top,
    colback=white,
    colframe=white,
    boxrule=0pt,
    arc=0pt,
    left=5pt,
    right=5pt,
    top=4pt,
    bottom=2pt,
    boxsep=0pt,
    drop fuzzy shadow=black!8
]

{\fontsize{7.1}{8.3}\selectfont

\textbf{Course Management}

\vspace{2pt}

\textbf{\color{subtext}
Purpose: 2 teachers \quad Rules: 0 teachers}

\vspace{2pt}

\color{bodytext}
Supports teacher-facing instructional preparation, lesson planning, and resource development.

\vspace{2pt}

\textbf{Examples from data:}

\begin{itemize}[
    leftmargin=9pt,
    itemsep=-0.5pt,
    topsep=0pt,
    parsep=0pt,
    partopsep=0pt
]

\item \textit{Generate lesson plans}
\item \textit{Create activities and assessments}

\end{itemize}

}

\end{tcolorbox}

\end{tabularx}

\end{tcolorbox}


\vspace{-2pt}

\caption{
Distribution of teacher-configured Purpose and Rule themes.
Each theme is shown once, with separate counts indicating the
number of teachers whose Purpose and Rule configurations reflected
that theme.
}

\Description{
A merged taxonomy of teacher-configured chatbot Purpose and Rule
themes. High-occurrence themes are Task/Objective, Pedagogical
Strategy, and Course Material. Medium-occurrence themes are
Constraints/Guardrails and Personalization. Low-occurrence themes
are Persona/Tone, Content Format, and Course Management. Separate
teacher counts are reported for Purpose and Rules within each theme.
}

\label{fig:purpose_rule_taxonomy}

\end{figure*}

\subsection{RQ3: Alignment Between Teachers' Envisioned Instructional Goals, Chatbot Configurations, and Generated Responses}

To address RQ3, we examined alignment across three stages: teachers'
instructional intentions expressed in the focus groups, how those intentions
were represented in their chatbot configurations, and how the resulting
chatbots behaved during interaction. We first report teachers' qualitative
observations of alignment and mismatch, followed by a log-based evaluation of
configuration alignment in generated responses. 

\subsubsection{Theme 7: Alignment and Gaps between Configured Intentions and Generated Behavior}
Approximately eight teachers described instances in which chatbot behavior
either reflected or diverged from their configured intentions. In some cases,
configured boundaries worked as intended. For example, a teacher in the
\T{13}--\T{15} focus group tested a water-cycle chatbot with both on-topic and
off-topic questions, explaining, ``I first asked a question about water cycle
and then [it] gave me a perfect answer...I also asked a question outside the
water cycle...and then it said, no, I can't do with this one.''

In other cases, configured intentions were not enacted consistently. \T{03} and \T{05} configured the chatbot to support
open-ended, step-by-step reasoning but observed that ``it doesn't seem like
it's meeting the responses'' they intended. \T{26} also found that adherence could vary by model: ``you put like don't give the answer, and then the one model gave the answer and the other didn't.'' These accounts show that expressing an instructional intention in the configuration did not always guarantee that the chatbot would enact it consistently.

\subsubsection{Log-Based Evaluation of Configuration Alignment in Chatbot Responses}
The final analysis included 27 unique chatbot IDs, with one final configuration
retained for each bot and duplicated configurations excluded. Using this
bot-level dataset, we examined alignment across the four evaluation dimensions.
The results suggest that the bots were generally responsive and aligned with
the configured persona, but showed less consistent alignment with the configured
purpose and rules (Table~\ref{tab:evaluation-results}). 

Responsiveness was the strongest dimension (88.9\%, Avg = 3.67), indicating that most bots were able to produce relevant and usable responses when users interacted with them. Persona also performed well (81.5\%, Avg = 3.48), suggesting that teachers were generally successful in configuring the chatbot’s role, tone, or identity in a way that was reflected in its responses.

Rules showed more moderate performance (70.4\%, Avg = 3.26). This means that although many configured behavioral constraints were followed, rule adherence was not fully reliable. Some bots may have responded appropriately overall while still violating or overlooking specific instructions established by the teacher.

Purpose showed the lowest alignment, with a 59.3\% pass rate and the lowest average score (Avg = 3.00). This indicates that a substantial proportion of the final chatbot cases did not meet the criterion for alignment with the instructional goal or intended function defined by the teacher. In other words, a chatbot could generate a reasonable response without consistently reflecting why the teacher created the bot in the first place.


Taken together, the pattern suggests a possible gap between general
conversational performance and fidelity to teacher-defined configurations. The
bots were more successful at being responsive and adopting the configured
persona than at consistently reflecting the configured instructional purpose
and rules in their responses. Purpose alignment therefore emerged as the
weakest dimension, followed by rule adherence.

\begin{table}[!t]
\centering
\caption{Pass rates and average scores for chatbot-response alignment across evaluation criteria.}
\label{tab:evaluation-results}
\Description{
Table reporting bot-level alignment results across four evaluation criteria:
Responsiveness, Purpose, Rules, and Persona. Responsiveness showed the strongest
performance, with 24 passes and 3 failures, an 88.9\% pass rate, and an average
score of 3.67. Persona followed with 22 passes, 5 failures, an 81.5\% pass rate,
and an average score of 3.48. Rules had 19 passes and 8 failures, corresponding
to a 70.4\% pass rate and an average score of 3.26. Purpose showed the lowest
alignment, with 16 passes and 11 failures, a 59.3\% pass rate, and an average
score of 3.00.
}
\begin{tabular}{lrrrr}
\toprule
\textbf{Metric} & \textbf{Pass} & \textbf{Fail} & \textbf{Pass \%} & \textbf{Avg Score} \\
\midrule
Responsiveness & 24 & 3  & 88.9\% & 3.67 \\
Purpose        & 16 & 11 & 59.3\% & 3.00 \\
Rules          & 19 & 8  & 70.4\% & 3.26 \\
Persona        & 22 & 5  & 81.5\% & 3.48 \\
\bottomrule
\end{tabular}
\Description{Evaluation results for responsiveness, purpose, rules, and persona, showing pass and fail counts, pass percentages, and average scores.}
\end{table}

\subsection{Theme 8 : Exploratory Findings: Making Student--AI Activity Visible and Actionable for Teachers}

Beyond the three research questions, teachers discussed how a teacher-facing dashboard could make students' interactions with AI more useful for classroom decision-making. Their comments reflected two related needs: understanding
students' activity and learning progress, and translating that visibility into
timely intervention without creating excessive monitoring or information
overload.

In \textit{Monitoring Student Activity, Understanding, and Progress}, teachers wanted visibility  beyond whether students were simply using the chatbot. They wanted to identify where students were in a learning process, who was
struggling, and what students appeared to understand. \T{07} described
wanting to see ``if they're ahead or if they're behind...what process of the
writing are they in...what step of the worksheet might they be in...who's
struggling?'' \T{09} noted that such information could show ``where my
students are at in their content knowledge'' and provide ``instant feedback.''
Together, these accounts position interaction data as a potential indicator of
learning needs rather than merely a record of chatbot use.

In \textit{Actionable Intervention, Selective Detail, and Privacy}, teachers
emphasized that this information should help them decide when attention was
needed without requiring review of every interaction. \T{06} explained
that ``the flag...would be our go-to...instead of having to check each
individual one,'' favoring high-priority alerts and real-time indicators.
Teachers also preferred summary-level information with the option to inspect
specific histories when an issue was flagged.

At the same time, increased visibility raised concerns about surveillance and
student privacy. \T{22} cautioned that parents might not want teachers
``watching what my kid's doing at all times.'' These tensions suggest that a
useful teacher dashboard should make student needs visible enough to support
action while limiting monitoring to information that is instructionally
relevant and necessary.

\subsection{Exploratory Patterns of Teacher Persona Combinations in Bot Personas}
\begin{figure*}[t]
    \centering
    \includegraphics[width=\textwidth]{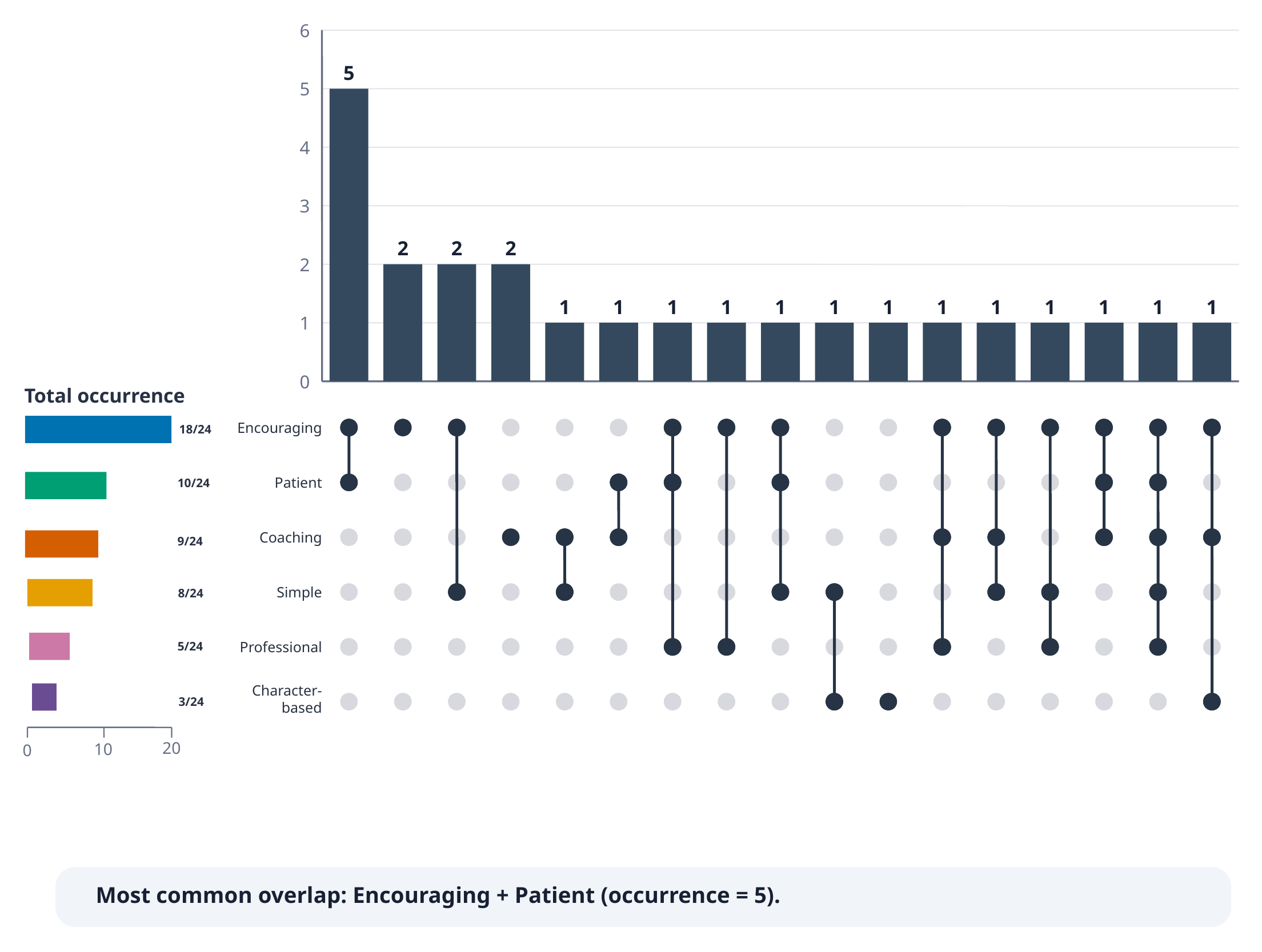}
    \caption{
    Exploratory analysis of persona and tone theme combinations across teachers' bot configurations. 
    The occurrence bars show the frequency of each persona theme, while the UpSet matrix shows the exact combinations of themes used together by teachers.
    }

  \Description{
UpSet plot showing the occurrence and co-occurrence of six persona themes across 24 teacher chatbot configurations. Encouraging is the most common theme, appearing in 18 configurations, followed by Patient in 10, Coaching in 9,Simple in 8, Professional in 5, and Character-based in 3. The most frequent exact combination is Encouraging with Patient, occurring in five configurations.
Three additional combinations occur twice each, while the remaining displayed
combinations occur once, indicating that teachers commonly combined persona
attributes rather than relying on a single characteristic.
}

    \label{fig:persona_themes}
\end{figure*}

Of the 27 teachers included in the configuration analysis, 24 specified at least one Persona attribute in their selected chatbot configuration. The remaining three teachers did not specify a persona or tone and were therefore excluded from the persona co-occurrence analysis.
Figure~\ref{fig:persona_themes} presents an exploratory analysis of how persona and tone themes were combined across teachers' bot configurations. The occurrence bars show that Encouraging was the most prevalent theme (18 occurrences), followed by Patient (10), Coaching (9), Simple (8), Professional (5), and Character-based (3). The UpSet matrix further shows that these themes were often used in combination rather than as isolated persona characteristics. The most common exact combination was Encouraging and Patient, appearing together for five teachers, while other configurations combined Encouraging with Simple, Coaching, or Professional traits. Overall, these patterns suggest that teachers did not treat persona as a single stylistic choice. Instead, they constructed composite bot personas by layering relational characteristics such as encouragement and patience with instructional roles such as coaching and communication preferences such as simplicity or professionalism.

\subsection{Operationalizing Personalization Across Bot Configuration Fields}

\begin{figure*}[!t]

    \centering

    \includegraphics[width=\textwidth]{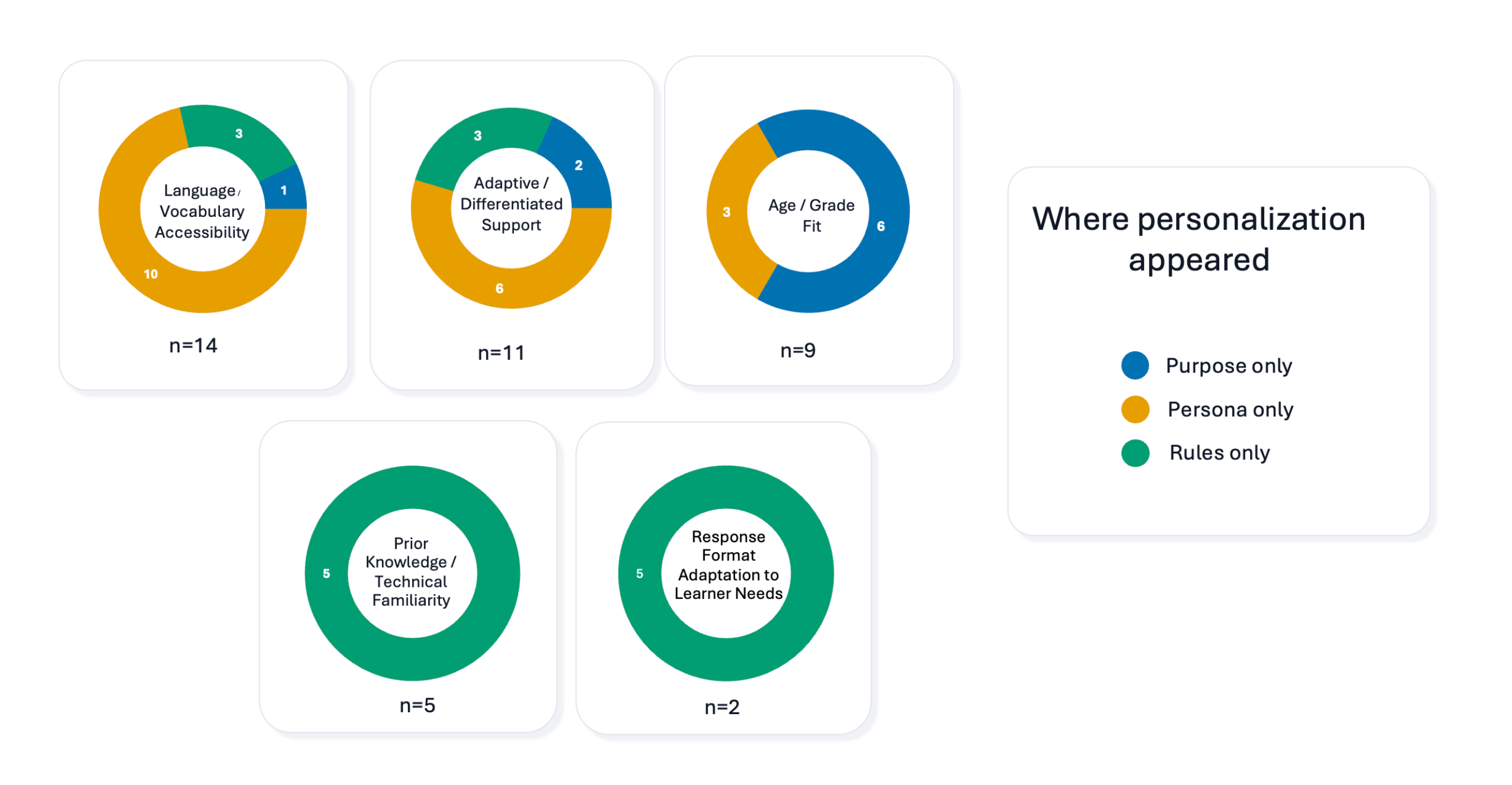}

    \caption{
    Distribution of personalization targets across Chatbot configuration fields.
    Each donut represents a personalization target, while segments indicate whether
    teachers expressed that target through Purpose, Persona, Rules fields.
    $n$ indicates
    the number of unique teachers associated with that target.
    }

    \Description{Five donut charts show how teachers expressed different personalization
    targets across chatbot configuration fields. The targets are language
    and vocabulary accessibility, adaptive or differentiated support, age or
    grade fit, prior knowledge or technical familiarity, and representation
    or modality adaptation. Each donut is divided into Purpose, Persona and Rules segments. Language and vocabulary accessibility and
    adaptive support are most often expressed through Persona, while prior
    knowledge and representation or modality adaptation are expressed through
    Rules. Age or grade fit is distributed across Purpose, Persona. Each chart also reports the number of unique teachers.}

    \label{fig:personalization_donut}

\end{figure*}

Figure~\ref{fig:personalization_donut} shows how teachers operationalized different forms of personalization across the Purpose, Persona, and Rules configuration fields. Overall, personalization appeared in the Purpose configurations of 9 teachers and in the Rules configurations of 13 teachers,
although teachers could express more than one type of personalization across their configurations.

Language and vocabulary accessibility was most often encoded through Persona, with 10 of 14 teachers expressing this form of personalization through Persona, compared with three through Rules and one through Purpose. A similar pattern appeared for adaptive or differentiated support, where 6 of 11 teachers used Persona, three used Rules, and two used Purpose. In contrast, assumptions about
students' prior knowledge or technical familiarity were expressed entirely through Rules (5 of 5 teachers), as were the two cases involving response-format adaptations to learner needs.

Age- and grade-level targeting ($n=9$) showed a different pattern: six teachers expressed it through Purpose and three through Persona. Overall, these patterns suggest that teachers treated personalization as a multidimensional authoring task, using different configuration fields to express different forms of
learner adaptation.

\section{Discussion}
\subsection{Contribution and Consistency with Prior Work }
We drew on the customization categories identified by Hou et al.~\cite{hou2026bespoke} as a starting point for our analysis. Hou et al. used these categories to examine instructors' customization priorities, grouping them into high-, medium-, and low-priority dimensions. Their findings showed that categories such as Pedagogical Strategy and Course Material were generally prioritized more highly, whereas Persona/Tone received lower priority.
We examined these same dimensions in chatbot configurations that teachers created themselves during the workshop. Rather than asking which customization dimensions teachers considered important, we analyzed where and how those pedagogical intentions were actually encoded within the authoring interface. For example, Task/Objective appeared primarily in the Purpose field, whereas Pedagogical Strategy and Course Material were expressed more often through Rules and Guidelines.
Our contribution therefore extends beyond identifying what teachers value in chatbot customization. We show how pedagogical intentions are operationalized through specific configuration fields and then examine whether those configurations are reflected in the chatbot's generated behavior as teachers intended. By combining configuration logs, teachers' testing messages, and the corresponding generated responses, our analysis traces the process from pedagogical intention, to authored configuration, to observed chatbot behavior.

\subsection{Teacher Control Through Configurable AI Authoring}

Our findings further show that teacher control over instructional AI depends on whether teachers can effectively express their pedagogical intentions through the available configuration options. Teachers used different fields for different functions: Purpose was used mainly to express the chatbot's instructional goal and content focus, while Rules were used more often to specify pedagogical behavior, guardrails, and learner-specific adaptations. This suggests that teachers benefit from distinct configuration mechanisms for different pedagogical functions, rather than a single general-purpose field.
However, configurability alone does not make authoring easy. Teachers sometimes struggled to interpret configuration options and requested clearer labels, examples, templates, and in-interface guidance. They also refined their configurations through testing and worked around interface constraints. This iterative process is consistent with prior work showing that teachers repeatedly test and refine pedagogical chatbots to better align generated responses with their instructional intentions~\cite{yoo2025teachers}. These tools should therefore support not only configuration, but also interpretation and refinement of how settings affect chatbot behavior.


\subsection{Bridging Pedagogical Intent and AI Behavior}

\begin{figure}[t]
    \centering
    \includegraphics[width=\linewidth]{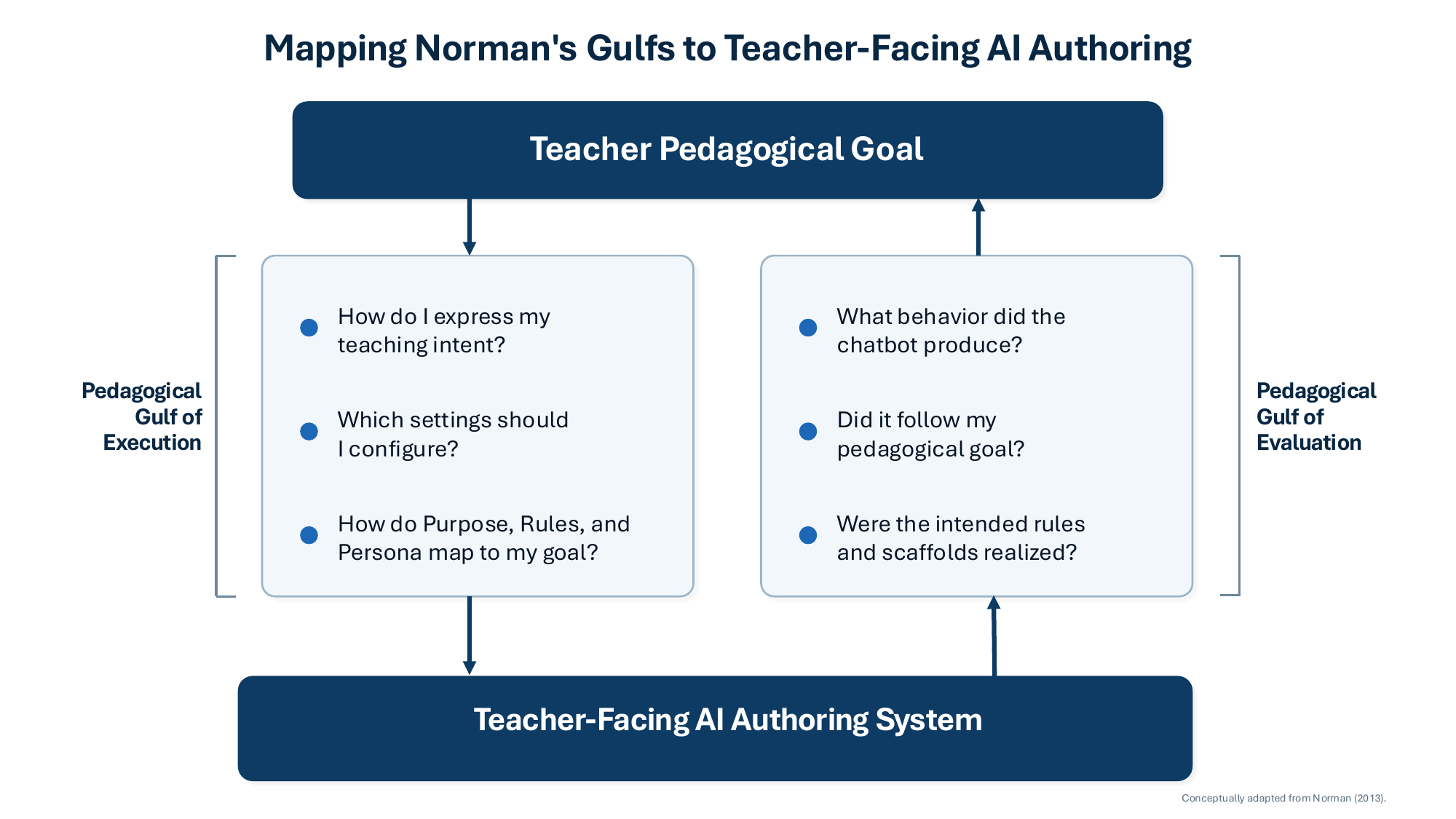}
    \caption{Mapping Norman's Gulfs of Execution and Evaluation to teacher-facing AI authoring. Adapted from Figure~2.1 in Norman~\cite{norman2013design}. In our adaptation, the pedagogical Gulf of Execution captures the translation of a teacher's pedagogical goal into chatbot configuration, while the pedagogical Gulf of Evaluation captures the interpretation of generated chatbot behavior relative to that original goal.}
   \Description{Conceptual diagram mapping Norman's Gulf of Execution and Gulf of Evaluation to teacher-facing AI authoring. A teacher pedagogical goal is shown at the top and a teacher-facing AI authoring system at the bottom. The left side represents the pedagogical Gulf of Execution, showing how teachers translate pedagogical goals into configuration choices such as Purpose, Rules, and Persona. The right side represents the pedagogical Gulf of Evaluation, showing how teachers interpret generated chatbot behavior and assess whether it reflects their original pedagogical goals.}
  \label{fig:pedagogical_gulfs}
\end{figure}

A conversationally appropriate response does not necessarily reflect the teacher's intended pedagogical behavior. Chatbot responses showed stronger alignment in responsiveness and persona than in purpose and rules with purpose emerging as the weakest dimension. This distinction suggests that evaluating educational AI requires considering not only conversational quality, but also fidelity to teacher-defined instructional goals and behavioral constraints.
Purpose may be more difficult to reflect consistently in individual chatbot responses because it often describes a broader instructional goal, whereas Rules and Persona provide more direct guidance about how the chatbot should respond. This may help explain why Purpose showed lower alignment than the
other dimensions.
We interpret these challenges through Norman's \textit{Gulf of Execution} and \textit{Gulf of Evaluation}. The Gulf of Execution describes the gap between a user's goal and the actions available to carry it out, while the Gulf of Evaluation describes the gap between a system's output and the user's ability to determine whether that output satisfies the original goal~\cite{norman1986cognitive}. We extend this framing to teacher-facing AI authoring by identifying pedagogical forms of both gulfs. Figure~\ref{fig:pedagogical_gulfs} adapts Norman's representation of these concepts to our context ~\cite{norman2013design}.

In our study, the \textit{pedagogical Gulf of Execution} captures the distance between a teacher's intended pedagogical behavior and the configuration actions available for expressing it. Teachers had to translate instructional intentions into fields such as Purpose, Rules, and Persona, a process that was not always straightforward.
The \textit{pedagogical Gulf of Evaluation} captures the distance between generated behavior and the teacher's ability to judge whether that behavior reflects the original pedagogical intention. This distinction is particularly important for generative AI, where a response may appear conversationally appropriate without fully realizing the intended instructional purpose.
Together, these two gulfs show why configurable controls alone are insufficient: teacher-facing AI authoring must support both the expression of pedagogical intentions and the evaluation of whether those intentions are realized in system behavior.

\subsection{Implications for Classroom AI Design}
Beyond these authoring and alignment challenges, the findings clarify the instructional roles teachers want configurable AI systems to support. Teachers envisioned chatbots as scaffolds that could provide differentiated support, help students work through tasks, and extend access to assistance when teachers were unavailable. At the same time, they wanted to preserve student thinking and authorship and maintain boundaries on what the chatbot could provide. These findings highlight the importance of supporting teacher-defined instructional boundaries and of examining how pedagogical intent carries from stated goals to configuration and generated behavior.

\section{Conclusion and Limitation}

This study contributes to research on educational AI by moving beyond the question of what teachers want to customize and examining how pedagogical intentions are translated into concrete chatbot configurations and whether those configurations are reflected in generated behavior. Our findings show
that different authoring fields served different pedagogical functions, while alignment was stronger for responsiveness and persona than for teacher-defined purpose and rules. This finding reinforces the need to evaluate educational AI not only in terms of response quality, but also in terms of fidelity to
educator-defined instructional goals. For the design of AI tools and educational chatbots, our results suggest the need for clearer guidance to help teachers translate instructional goals into configuration settings, as well as mechanisms for testing, diagnosing, and refining chatbot behavior before classroom deployment. Such support is particularly important in K--12 education, where maintaining teacher agency requires educators to retain meaningful control over how AI scaffolds learning, establishes instructional boundaries, and interacts with students.

Our findings also open several directions for future research. For example, future AI authoring systems could provide automated feedback indicating which parts of a teacher's configuration are not strongly reflected in generated responses, recommend revisions to Purpose or Rules, and allow teachers to
compare how different language models enact the same configuration. Classroom studies could further examine how teachers revise their configurations after observing authentic student interactions and whether greater configuration fidelity leads to more effective and pedagogically appropriate support.

A key limitation of this study is that chatbot behavior was evaluated primarily through teachers' testing interactions during relatively short professional development workshops. Although introducing teachers to the platform and providing initial hands-on practice was necessary before they could meaningfully configure and evaluate their chatbots, this setting does not capture sustained student use in authentic classrooms. Consequently, the observed alignment may not reflect failures or adaptations that emerge during longer, more varied, or unexpected student interactions.

\FloatBarrier
\bibliographystyle{ACM-Reference-Format}
\bibliography{reference}

\appendix

\section{Pre-Survey}
\label{app:presurvey}

\includepdf[
    pages=-,
    pagecommand={},
    width=\textwidth
]{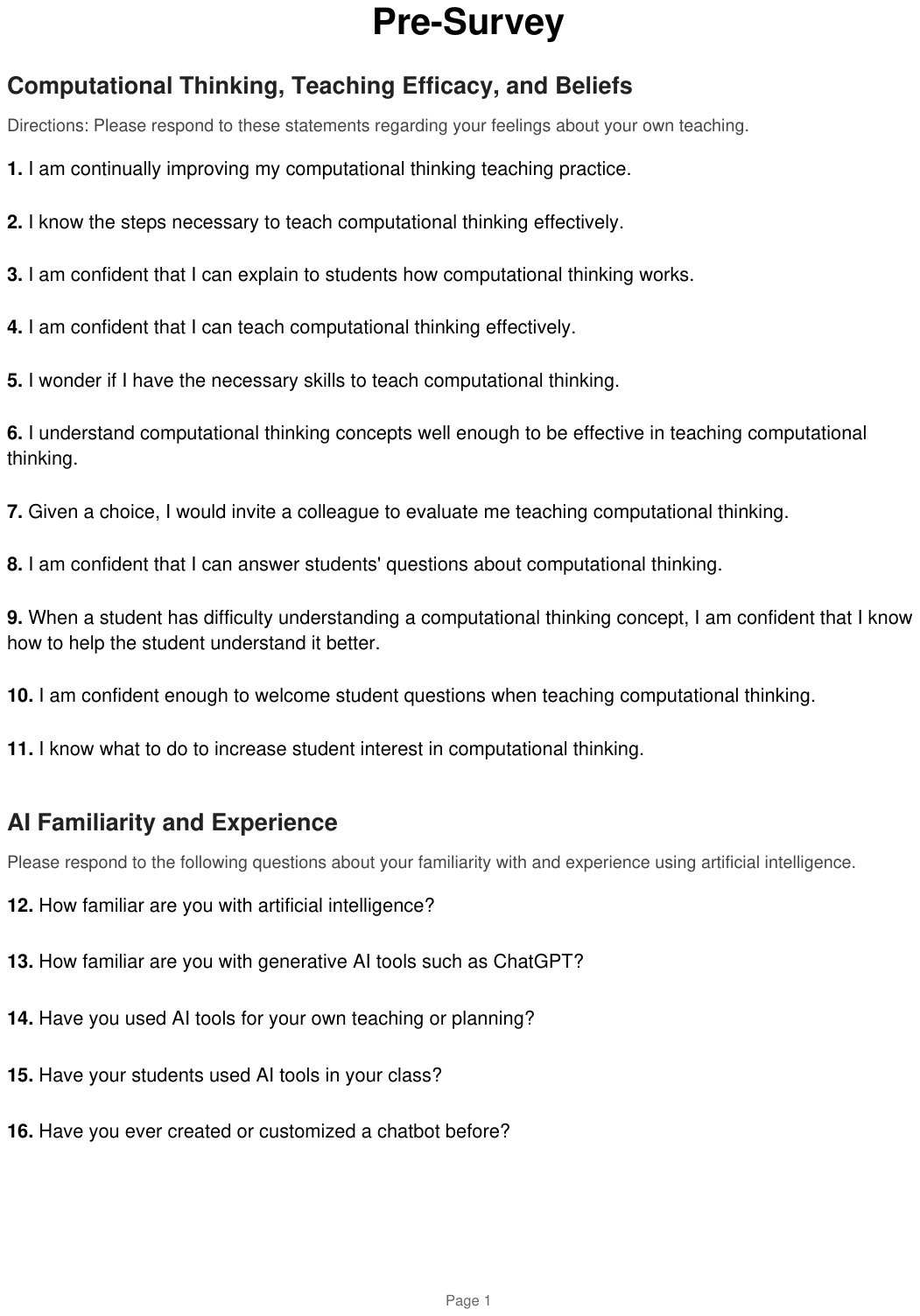}


\end{document}